\documentclass{pasa}

\title[Mopra CO Data Release 1]{The Mopra Southern Galactic Plane CO Survey --- \\Data Release 1}
\author[Braiding et al.]{Catherine Braiding$^{1,12}$, M.~G. Burton$^1$, R. 
Blackwell$^2$, C. Gl\"uck$^3$, J. Hawkes$^2$, C. Kulesa$^4$, N. Maxted$^5$, \\
D. Rebolledo$^{1,6}$, G. Rowell$^2$, A. Stark$^7$, N. Tothill$^8$, J.~S. Urquhart$^9$, 
F. Voisin$^2$, A.~J. Walsh$^{10}$, P. de Wilt$^2$\\
and G.~F. Wong$^{8,11}$\\
    \affil{$^1$School of Physics, University of New South Wales, Sydney, NSW 2052, Australia}
    \affil{$^2$Department of Physics, School of Physical Sciences, University of Adelaide, Adelaide, SA 5005, Australia}
    \affil{$^3$KOSMA, I. Physikalisches Institut, Universit\"at zu K\"oln, Z\"ulpicher Str. 77, 50937 K\"oln, Germany}
    \affil{$^4$Steward Observatory, The University of Arizona, 933 N. Cherry Ave., Tucson, AZ 85721, USA}
    \affil{$^5$Laboratoire Univers et Particules de Montpellier, Universite de Montpellier 2, Montpellier, Herault 34000, France}
    \affil{$^6$Sydney Institute for Astronomy, University of Sydney, Sydney, NSW 2006, Australia}
    \affil{$^7$Harvard-Smithsonian Center for Astrophysics, 60 Garden Street, Cambridge, MA 02138, USA}
    \affil{$^8$School of Computing Engineering and Mathematics, University of Western Sydney, Locked Bay 1797, Penrith, NSW 2751, Australia}
    \affil{$^9$Max-Planck-Institut f\"ur Radioastronomie, Auf dem H\"ugel 69, D-53121 Bonn, Germany}
    \affil{$^{10}$International Centre for Radio Astronomy Research, Curtin University, GPO Box U1987, Perth, WA 6845, Australia}
    \affil{$^{11}$CSIRO Astronomy and Space Science, PO Box 76, Epping, NSW 1710, Australia}
    \affil{$^{12}$Email: catherine.braiding@gmail.com}}%
\jid{PASA}
\doi{10.1017/pasa.\the\year.xxx}
\jyear{\the\year}

\usepackage[authoryear]{natbib}
\bibpunct{(}{)}{;}{a}{}{,}
\usepackage{amsmath}
\usepackage{amssymb}
\begin{document}%
\begin{abstract}
We present observations of the first ten degrees of longitude in the Mopra carbon 
monoxide (CO) survey of the southern Galactic plane \citep{betal2013}, covering 
Galactic longitude $l = 320$--$330^\circ$ and latitude $b = \pm 0.5^\circ$, and 
$l = 327$--$330^\circ$, $b = +0.5$--$1.0^\circ$. These data have been taken at 
35 arcsec spatial resolution and 0.1\,km\,s$^{-1}$ spectral resolution, providing 
an unprecedented view of the molecular clouds and gas of the southern Galactic 
plane in the 109--115\,GHz $J = 1$--0 transitions of $^{12}$CO, $^{13}$CO, 
C$^{18}$O and C$^{17}$O. Together with information about the noise statistics from 
the Mopra telescope, these data can be retrieved from the Mopra CO website and the 
CSIRO-ATNF data archive.
\end{abstract}
\maketitle%
\section{INTRODUCTION }
\label{sec:intro}

Molecular clouds comprise the densest regions of the interstellar medium, and are the nurseries 
in which stars form. In order to determine how molecular clouds themselves form, surveys of the 
three principal forms of carbon (i.e.\ C$^+$, C and CO) are needed to trace the gas in its 
ionised, atomic and molecular forms respectively. In particular, surveys along the Galactic 
plane, where the great majority of molecular clouds lie, provide us with a picture of the 
molecular structure of our Galaxy.

The Mopra southern Galactic plane CO Survey is designed to map the distribution and dynamics of 
the carbon monoxide (CO) molecule along the Fourth Quadrant of the Galaxy, from $l = 
270$--$360^\circ$ with $b = \pm 0.5^\circ$ \citep[][Paper I]{betal2013}. Using the Mopra radio 
telescope, the $J$ = 1--0 line of CO at 2.6\,mm is being mapped at 35\,arcsec spatial and 
0.1\,km\,s$^{-1}$ spectral resolution, in order to determine where giant molecular clouds (GMCs) 
lie in the Galaxy and to explore the connection between molecular clouds and the `missing' gas 
inferred from gamma-ray observations. The CO data will be combined with similar survey data 
collected of the sub-millimetre neutral carbon line (C) taken with the Nanten2 telescope in 
Chile, and the terahertz neutral and ionized carbon lines (C and C$^+$) from the balloon-borne 
Stratospheric Terahertz Observatory 2 (STO2) and the ice-bound High Elevation Antarctic 
Telescope (HEAT), both in Antarctica. With these surveys it will be possible to determine the 
distribution and motion of the principal forms of carbon in the interstellar medium. 

\begin{figure*}
\includegraphics[width=\textwidth]{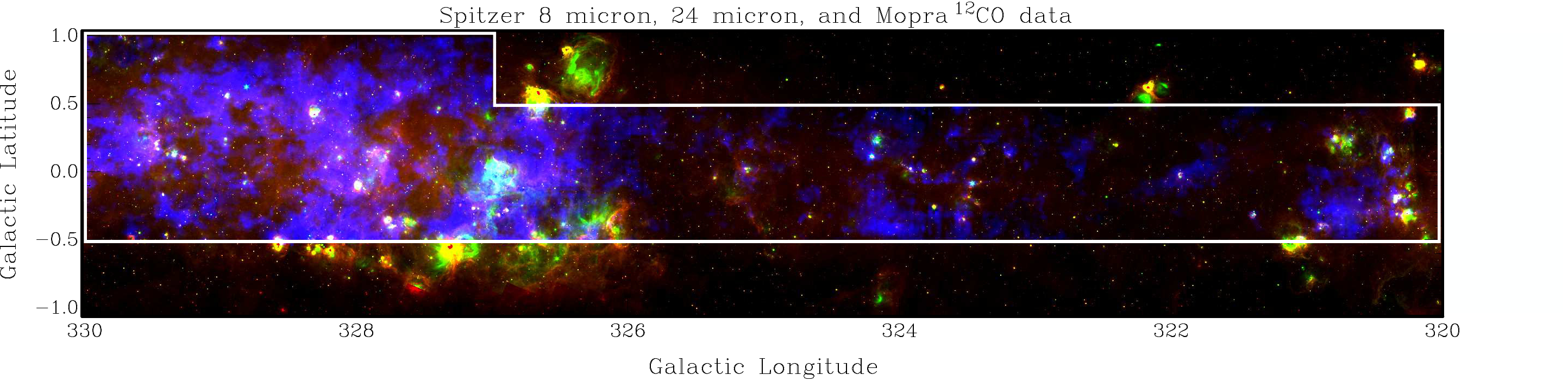}
\caption{Three-colour image of the Galactic plane from the Spitzer GLIMPSE 8\,$\mu$m 
\citep[red;][]{churchwell2009,betal2003} and MIPSGAL 24\,$\mu$m \citep[green;][]{carey2009} 
surveys, with a moment 0 map calculated over $v = -120$ to $0$\,km\,s$^{-1}$ from the Mopra 
$^{12}$CO $J=1$--$0$ survey (blue). The region enclosed by the white box, containing Galactic 
longitudes $l = 320$--$330^\circ$ and latitudes $b = \pm 0.5^\circ$, as well as 
$l=327$--$330^\circ$, $b = +0.5$--$1.0^\circ$, is that published here as Data Release 1 (DR1). 
\label{Fig1-map}}
\end{figure*}

\begin{figure*}
\includegraphics[width=\textwidth]{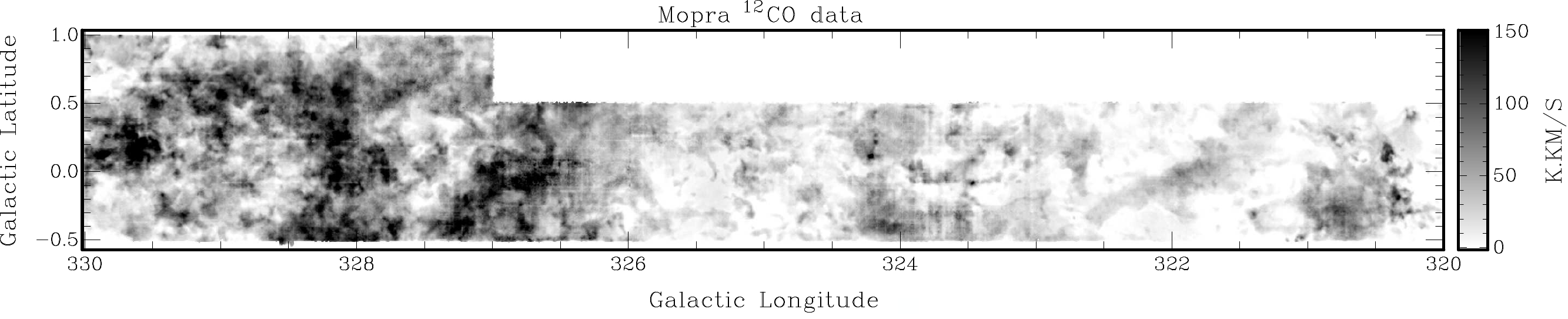}
\caption{The moment 0 map calculated over $v = -120$ to $0$\,km\,s$^{-1}$ from the Mopra 
$^{12}$CO $J=1$--$0$ Southern Galactic Plane survey. The region illustrated is that published 
here as Data Release 1 (DR1). \label{Fig1-map2}}
\end{figure*}

Data Release 1 (hereafter DR1) covers the first ten square degrees of the survey, from $l = 
320$--$330^\circ$ with $b=\pm0.5^\circ$, along with an additional three half-square degrees over 
$l = 327$--$330^\circ$, $b = +0.5$--$1.0^\circ$ (see Figures \ref{Fig1-map} \& \ref{Fig1-map2}). 
The G323 sightline has previously been published in Paper I, however, it is discussed here 
because the G323 data was taken in the summer months and compares less favourably to the other 
data cubes, which were obtained in the Austral winter. The improved atmospheric conditions make 
it possible to observe all four of the major CO isotopologues ($^{12}$CO, $^{13}$CO, C$^{18}$O 
and C$^{17}$O), although the C$^{17}$O line is very weak and only detectable at a $2\sigma$ 
level at best, as measuring it typically requires longer integration times than our fast-mapping 
survey mode allows (see Paper I). The isotopologue ratios allow examination of the distribution 
of columns of molecular gas through GMCs \citep{hetal2009}, so that variance in the X$_{\rm CO}$ 
factor, used to link the CO intensity to the H$_2$ column density in order to obtain molecular 
cloud masses, can be determined \citep[see e.g.][]{bwl2013}.

In Section 2 of this paper (Paper II) the data are described, along with the various metrics used 
to quantify the noise and data quality. Section 3 will present some sample science results, 
however detailed science investigations are left for focussed studies, such as \citet{betal2014}, 
which described a filamentary molecular cloud in the G328 sightline. The data are being made 
available at the survey website\footnote{http://www.phys.unsw.edu.au/mopraco/}, and in the 
CSIRO-ATNF data archive\footnote{http://atoa.atnf.csiro.au}, as they are published. 

\section{DATA RELEASE 1}

The data presented in this paper were taken in March 2011, and over the Austral winters of 
2011--2013. The observations performed in 2011 ($l = 323$--$330^\circ$, $b = \pm0.5^\circ$) used 
the UNSW Digital Filter Bank (the UNSW--MOPS) in its `zoom' mode, with $4\times137.5$\,MHz 
dual-polarization bands; in 2012 this was increased to 8 bands, so that the data from $l = 
320$--$323^\circ$, $b=\pm0.5^\circ$ and $l=327$--$330^\circ$, $b = +0.5$--$1.0^\circ$ cover a 
larger spectral region. Very little emission is observed in these extended bands away from the 
Galactic centre, but it is possible that high velocity clouds could be detected. The full line 
parameters are given in Table 1 of Paper I; the DR1 dataset has a spectral resolution of $\sim 
0.1$\,km\,s$^{-1}$ over at least 4096 channels for each of the four CO isotopologues.

The angular resolution of the Mopra beam at 115\,GHz is 33\,arcsec FWHM \citep{letal2005}; 
after the median filter convolution is applied during the data reduction process, the beam 
size is around 35\,arcsec in the final data set. The extended beam efficiency, used to 
convert brightness temperatures into line fluxes, is $\eta_{\rm XB} = 0.55$ at 115\,GHz 
\citep[rather than the main beam efficiency $\eta_{\rm MB} = 0.42$;][]{letal2005}. 

\begin{figure}
\begin{center}
\includegraphics[width=\columnwidth]{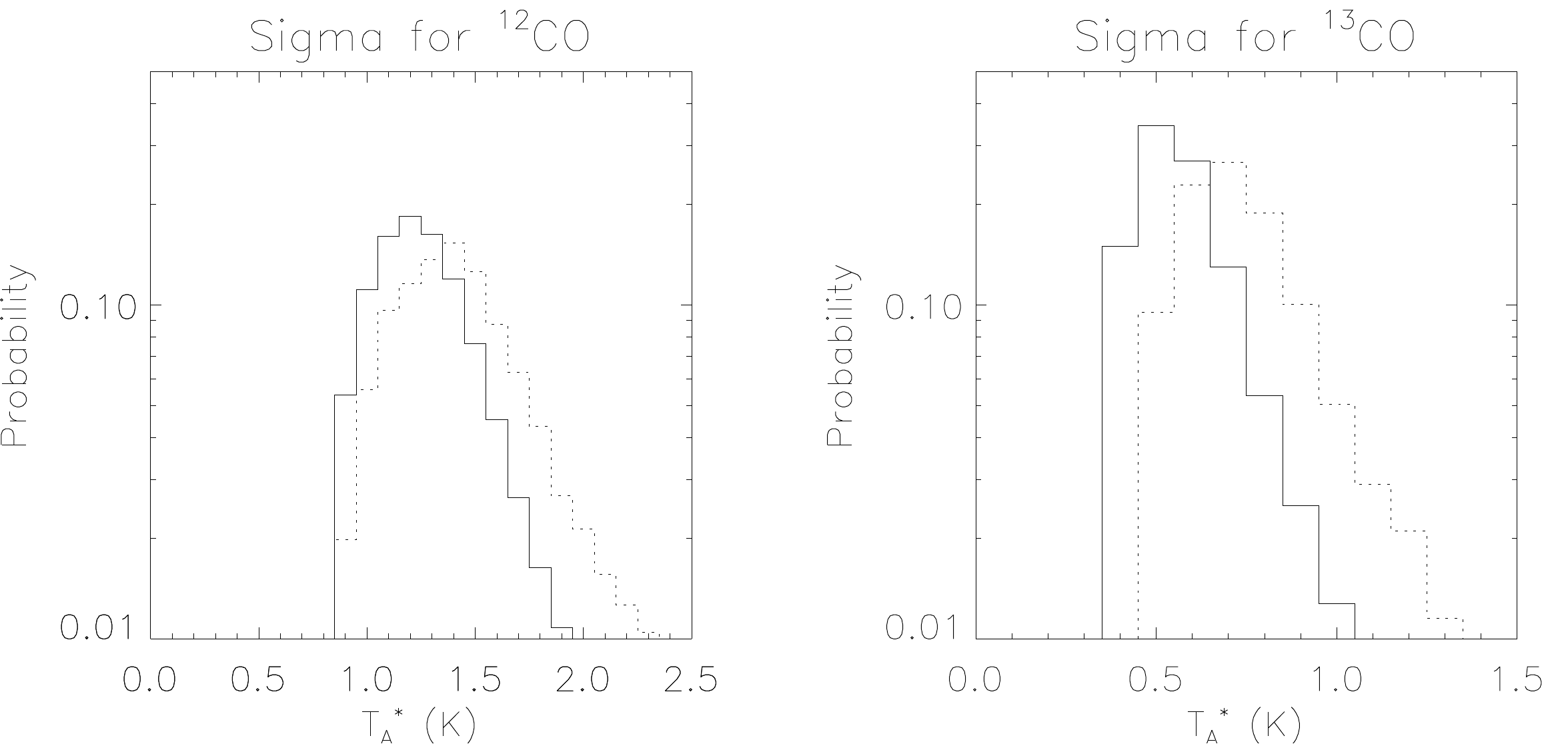}
\caption{Probability distribution of the noise level in the $^{12}$CO (left) and $^{13}$CO 
data cubes, $\sigma_{\rm cont}$, as determined from the standard deviation in the continuum 
channels (in $T_A^\star$\,[K] units) for each pixel. The dotted distribution in each is the 
noise level in the $l=323$--324$^\circ$ field (described in Paper I), highlighting the 
effect of observing in the summer months.\label{Sigma}}
\vspace{-10pt}
\end{center}
\end{figure}
\begin{figure*}
\begin{center}
\includegraphics[width=\textwidth]{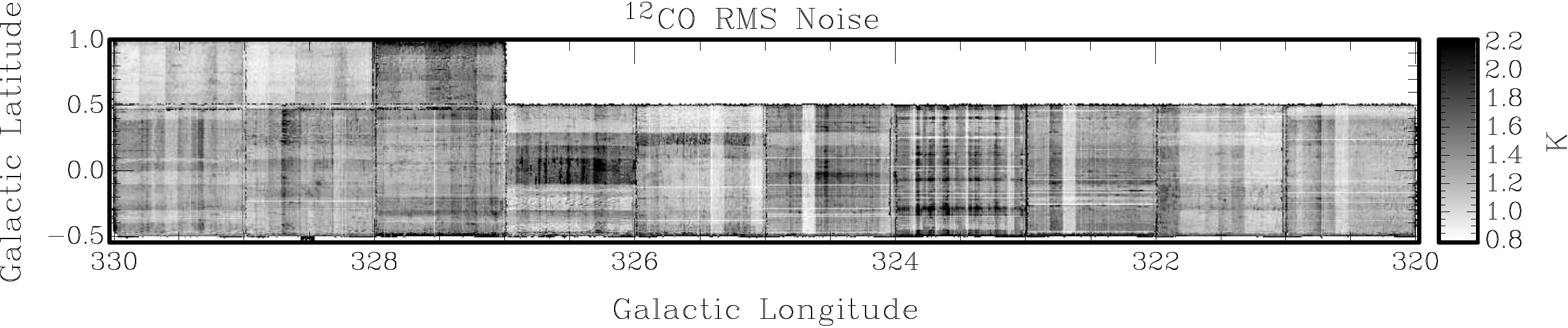}
\includegraphics[width=\textwidth]{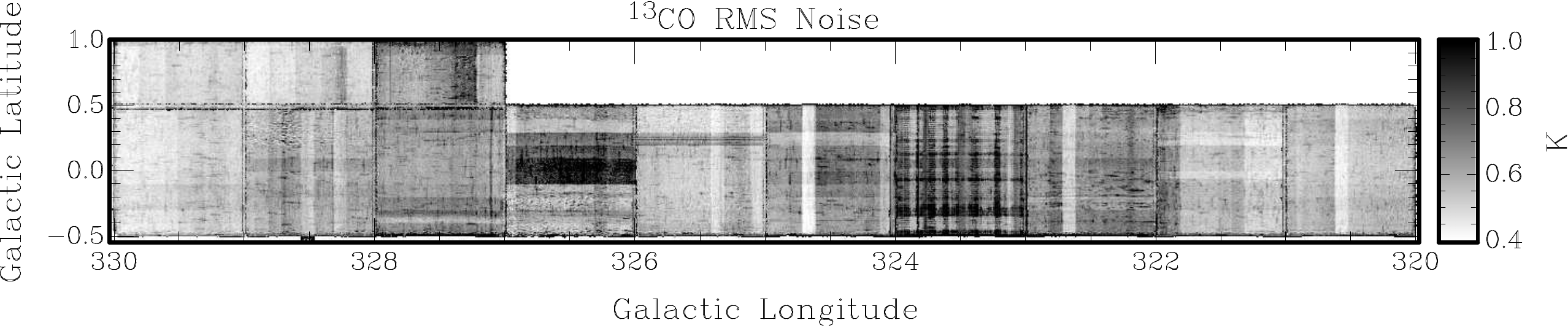}
\caption{Maps showing the noise level (in $T_A^\star$\,[K] units) for $^{12}$CO and $^{13}$CO, 
determined from the standard deviation of the continuum channels between 0 and 
$+90$\,km\,s$^{-1}$ for each pixel. Note that the intensity scales differ, as the $^{12}$CO 
observations have $\sim3\times$ higher noise than $^{13}$CO. The higher noise in the G323 region 
(Paper I) is also evident.\label{Sigma-map}}
\end{center}
\end{figure*}
There are a number of metrics used to assess the quality of DR1. Figure \ref{Sigma} shows 
histograms of the noise per channel, $\sigma_{\rm cont}$, as well as the previously-published 
distribution from the G323 sightline for both the $^{12}$CO and $^{13}$CO lines (the C$^{18}$O 
and C$^{17}$O lines have similar noise characteristics to $^{13}$CO). These are calculated by 
finding the standard deviation of the channels outside of the range where line emission 
occurs, where $V_{\rm LSR} < -120$\,km\,s$^{-1}$ and $V_{\rm LSR} > +20$\,km\,s$^{-1}$ (Figure 
\ref{Fig7c-Moments} shows that some emission exists where $V_{\rm LSR} > +20$\,km\,s$^{-1}$, 
but it occurs in very few pixels and does not affect these statistics). Although the G323 data 
were taken in the summer months, the noise distribution is not significantly different from 
those data obtained in winter. While the tails of the histograms are $\sim0.5$\,K higher in the 
summer observations due to the poorer conditions, the mode values for $^{12}$CO and $^{13}$CO 
are 1.5 and 0.7\,K per 0.1\,km\,s$^{-1}$ velocity channel, in comparison to 1.3 and 0.5\,K per 
0.1\,km\,s$^{-1}$ channel in winter. The $^{12}$CO line has higher noise than the $^{13}$CO line 
(and those of the other isotopologues not shown here), as there is a molecular oxygen absorption 
line near 115\,GHz, making the atmosphere inherently worse than it is for the other 
isotopologues at 109--112\,GHz. 

Figure \ref{Sigma-map} shows the $1\sigma$ noise maps for the $^{12}$CO and $^{13}$CO data 
cubes, determined from the standard deviation of the continuum channels between $v = 0$ and 
$v = +90$\,km\,s$^{-1}$ in each pixel. The wider visible striping in these maps is an effect 
of the scans in the $l$ and $b$ directions being taken with a variety of different system 
temperatures and observing conditions. Thinner stripes or spots of lower apparent noise occur 
where `bad' pixels, rows or columns were identified and interpolated over (the lower noise is 
a result of interpolating over several nearby pixels; see Paper I). As in Figure \ref{Sigma}, 
the higher noise occurring in the G323 field is a result of poorer conditions in the summer 
months.

\begin{figure}
\begin{center}
\includegraphics[width=\columnwidth]{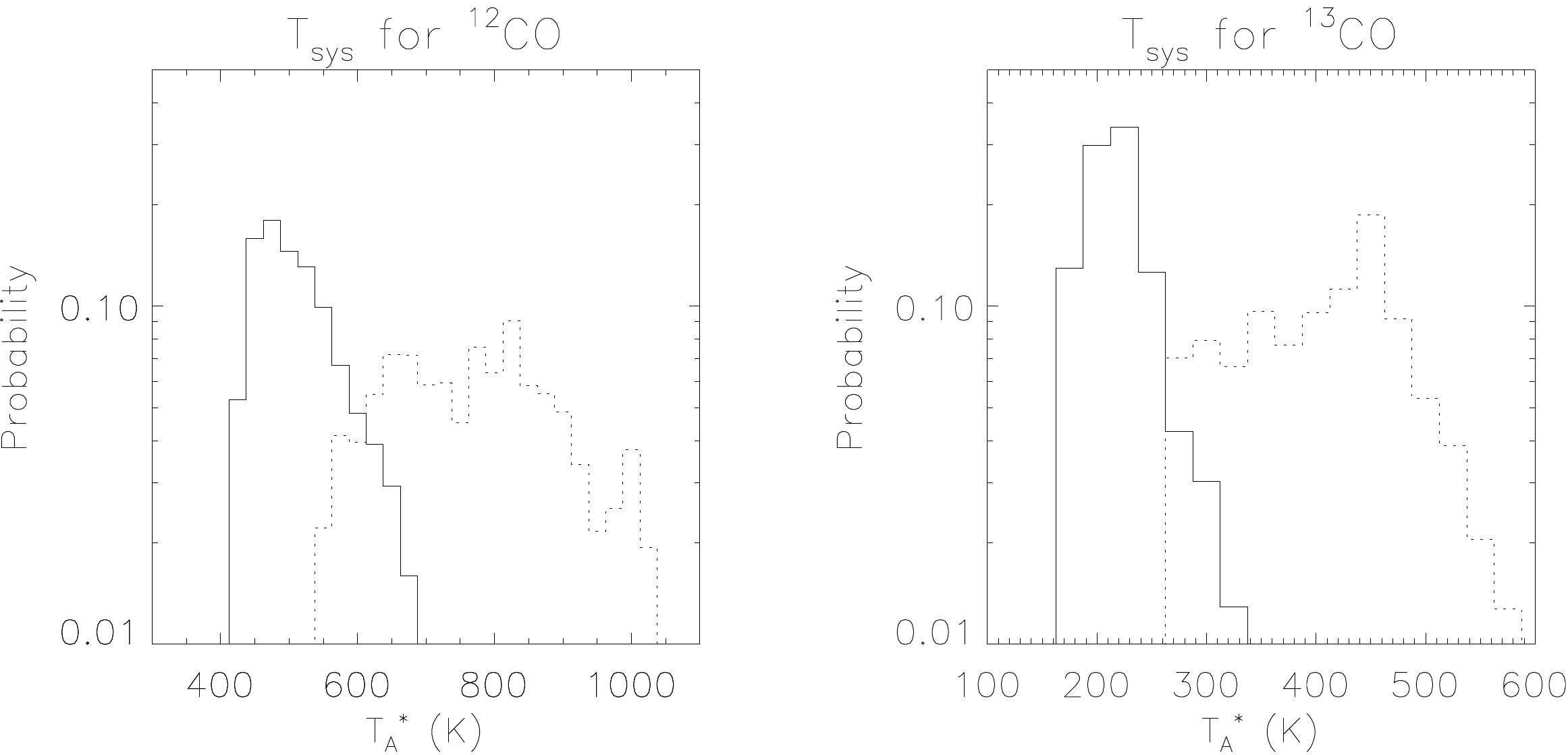}
\caption{Probability distribution of the system temperature, $T_{\rm sys}$ (in $T_A^\star$ [K] 
units) for $^{12}$CO and $^{13}$CO, in the data for each pixel, determined from the ambient 
temperature load paddle measurements. The dotted distribution in each is the $l = 
323$--324$^\circ$ field described in Paper I; the higher $T_{\rm sys}$ values highlight the 
effect of observing in the summer months.\label{tsys}}
\end{center}
\end{figure}
\begin{figure*}
\begin{center}
\includegraphics[width=\textwidth]{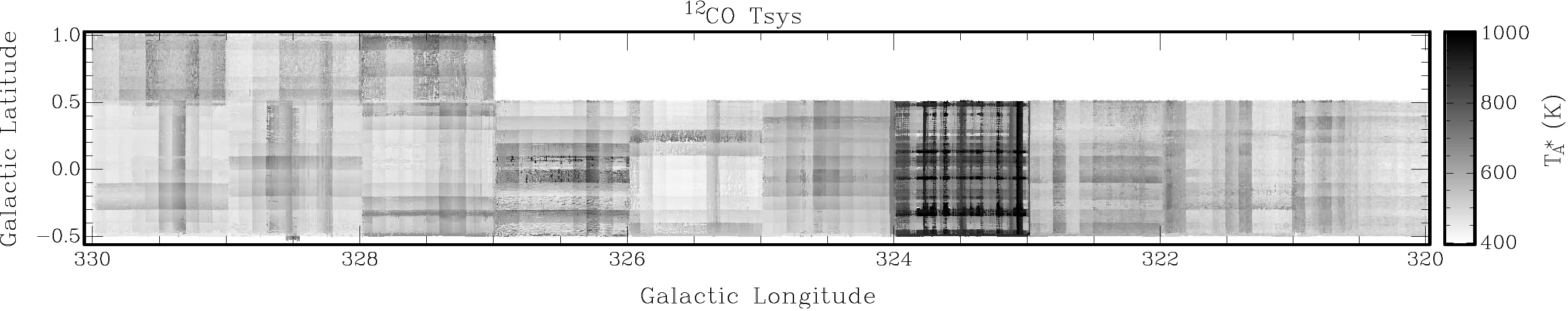}
\includegraphics[width=\textwidth]{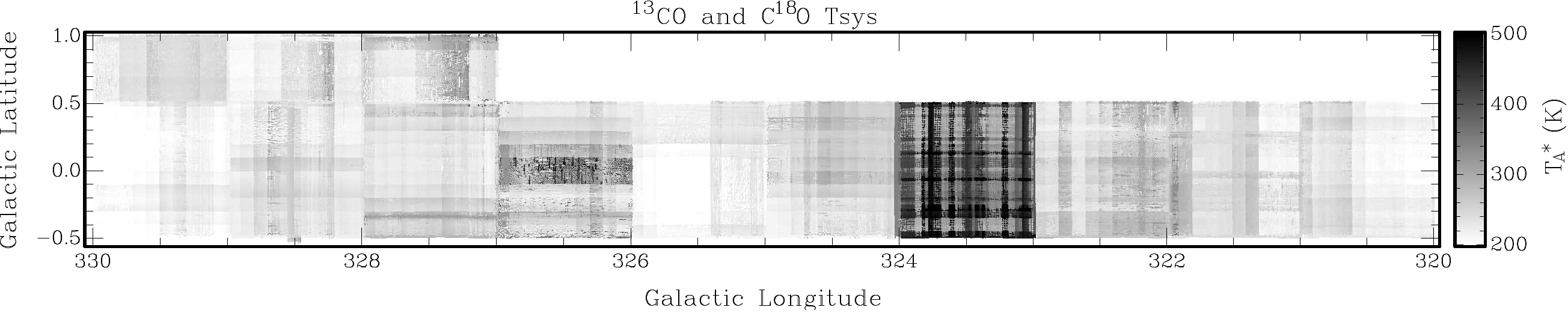}
\includegraphics[width=\textwidth]{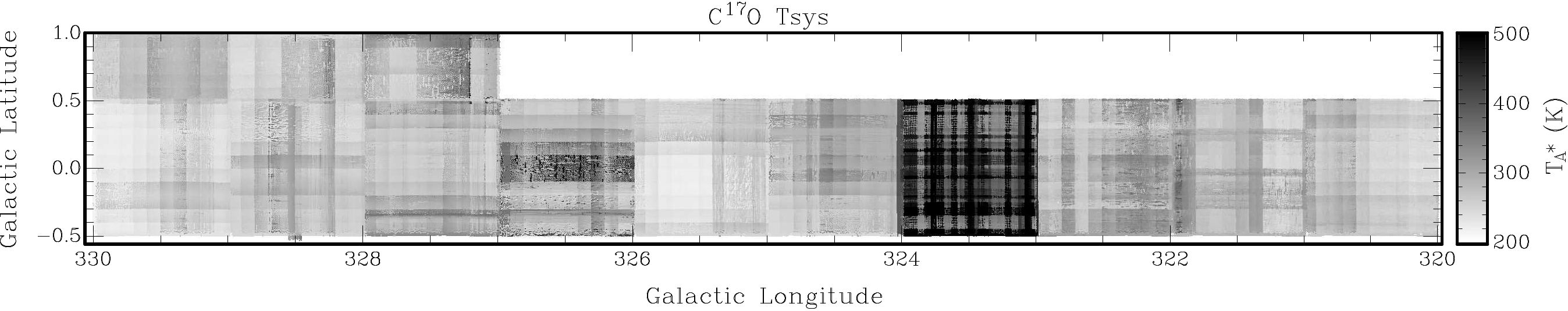}
\caption{$T_{\rm sys}$ images for (from top): $^{12}$CO, $^{13}$CO and C$^{18}$O (which share 
the same 2\,GHz band of the correlator), and C$^{17}$O, in units of $T_A^\star$\,(K) (as 
indicated by the scale bars). The striping pattern is inherent to the data set, resulting from 
scanning in the $l$ and $b$ directions in variable observing conditions. The darkest square 
degree region, $l = 323$--324$^\circ$, was observed in March 2011 when the summer conditions 
were less favourable for CO observations. 
\label{Fig4b-Tsys-map}}
\end{center}
\end{figure*}

The system temperature, $T_{\rm sys}$, measures the level of the received signal from the 
source, sky, telescope and instrument. It is calibrated every 30 minutes with reference to an 
ambient temperature paddle that is placed in front of the beam. Figure \ref{tsys} shows 
histograms of the probability distribution of $T_{\rm sys}$ for $^{12}$CO and $^{13}$CO, 
alongside their counterparts for the G323 sightline. Again, the $^{12}$CO line has 
$T_{\rm sys}$ about a factor of two higher than the values for the other lines due to the 
inferior conditions. Maps of the system temperature for all four of the CO isotopologues are 
shown in Figure \ref{Fig4b-Tsys-map}; $^{13}$CO and C$^{18}$O share the same $T_{\rm sys}$ map 
as these lines are in the same 2\,GHz band of the correlator. As before, the striping is a 
result of combining data from scans in both the $l$ and $b$ directions (minimizing artefacts 
from poor sky conditions), while spots occur where the data cube has been thresholded to 
remove that data with excessive $T_{\rm sys}$ values (see Paper I for more details). Scans 
with particularly high $T_{\rm sys}$ values were repeated (these are also visible in Figure 
\ref{Sigma-map} as wide stripes of lower noise).

Several of the data cubes were affected by contamination from faint molecular clouds in the 
sky reference beam (of typical intensity $\sim -0.5$\,K in $^{12}$CO and $\sim -0.1$\,K in 
$^{13}$CO; the other lines are generally too weak for sky emission to be visible). While tests 
of the sky reference positions were carried out prior to survey observations, these were quick 
observations and the faint sky signal was typically lost in the noise. In DR1 this 
contamination has been removed by calculating the average spectra in a region of $\sim 100$ 
pixels where the sky contamination is uniform, such that there is no true emission within 
$\pm 3$\,km\,s$^{-1}$ of the negative emission profile. A Gaussian template is fit to this 
contaminating profile and then subtracted from the individual spectrum at each pixel in the 
full square degree. The central velocity of the sky emission varies across each square degree, 
as the reduction process (see Paper I) does not control for any variation relative to the local 
standard of rest in the sky spectrum; this is compensated for by linearly varying the central 
velocity of the template spectrum in both longitude and latitude (typically by $\lesssim 1$\,km 
\,s$^{-1}$ in either direction). Table \ref{tab-sky} gives the position of the sky reference 
beam for each square degree, along with the intensity and central velocity of the contamination 
in the average spectrum for the square degree.  
\begin{table*}
\caption{Sky reference beam positions for each square degree of DR1; the additional half-square 
degree regions from $l=327$--$330^\circ$, $b = +0.5$--$1.0^\circ$ use the same sky reference 
positions as their main survey counterparts. Several of the cubes were affected by contamination 
due to $^{12}$CO line emission at the sky reference position; the central velocity and intensity 
(in $T_A^\star$\,[K] units) of this contamination is also noted.}
\begin{center}
\begin{tabular}{@{}ccccc@{}}
\hline\hline
Cube & Reference & Reference & Contamination          & Contamination Peak\\
     & Longitude & Latitude  & Velocity (km\,s$^{-1}$) & Intensity ($T_A^\star$\,[K]) \\
\hline%
 G320  & 320.375 & -3.000    \\
 G321  & 321.500 & -2.500    & -1.68                  & -3.18 \\
 G322  & 322.500 & -2.000    & -56.11, -12.96         & -1.25, -0.39 \\ 
 G323  & 323.500 & -2.000    \\ 
 G324  & 324.500 & -2.000    & +0.90                  & -0.78 \\ 
 G325  & 325.500 & -2.200    & -46.62                 & -0.49 \\ 
 G326  & 326.500 & -2.000    & -19.31                 & -0.55 \\ 
 G327  & 327.500 & -2.200    \\ 
 G328  & 328.500 & -2.200    & +2.68                  & -0.42 \\
 G329  & 329.500 & -2.200    \\ 
\hline\hline
\end{tabular}
\end{center}
\label{tab-sky}
\end{table*}

DR1 has been made available at the survey website and in the CSIRO-ATNF data archive as a 
series of fits files for each isotopologue, each containing a square degree of the survey, 
centred at each half degree along the Galactic plane. Although the full range of velocity 
space available for $^{12}$CO is $[-1000,+1000]$\,km\,s$^{-1}$ (with file sizes $\sim800$\,MB / 
square degree), it is expected that most users would prefer to download the smaller ($\sim 
300$\,MB / square degree) files covering the velocity range $[-150,+50]$\,km\,s$^{-1}$ 
illustrated in this paper. No emission has yet been detected outside this range, although 
closer to the Galactic centre a wider velocity range will be included within the smaller files. 
Also available are the $T_\textrm{sys}$, beam coverage and rms noise images for each line. The 
$10^\circ \times 1.5^\circ$ cube used for the analysis in this paper was created using the 
\texttt{imcomb} task in miriad\footnote{http://www.atnf.csiro.au/computing/software/miriad/} 
\citep{stw1995}.

\section{RESULTS}

\begin{figure*}
\begin{center}
\includegraphics[height=\textwidth,angle=90]{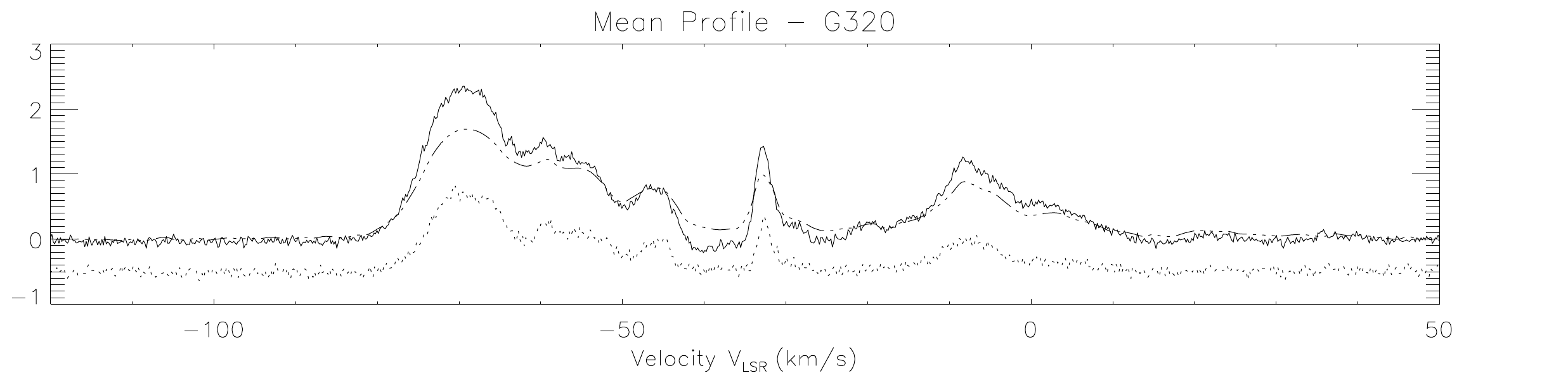}
\vspace{-3pt}
\includegraphics[height=\textwidth,angle=90]{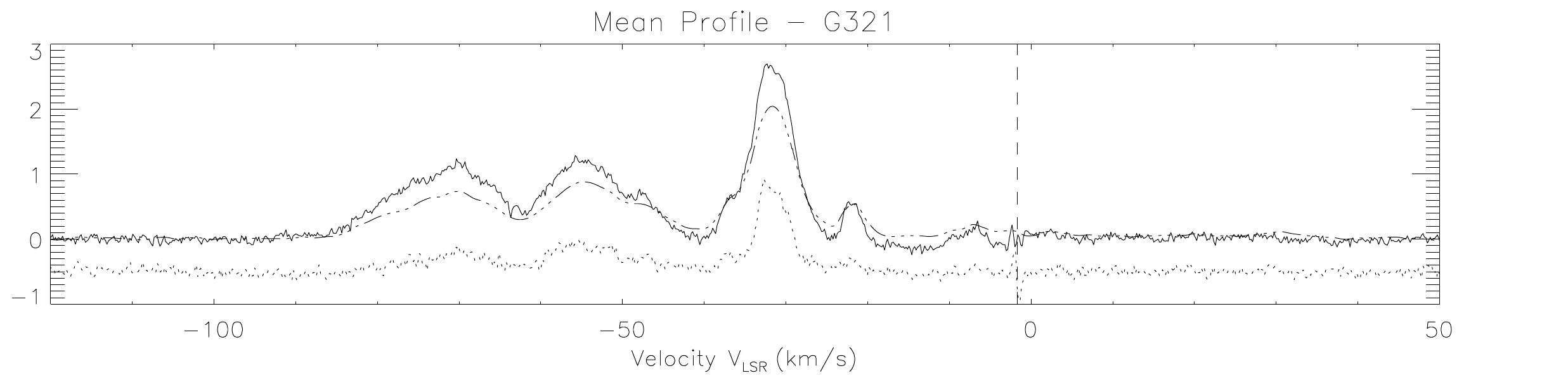}
\vspace{-3pt}
\includegraphics[height=\textwidth,angle=90]{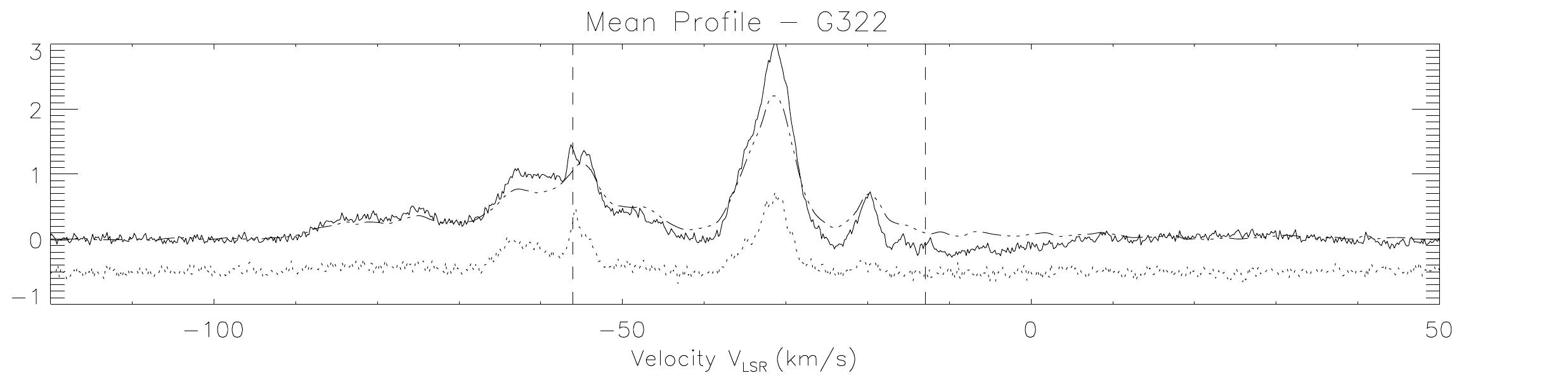}
\vspace{-3pt}
\includegraphics[height=\textwidth,angle=90]{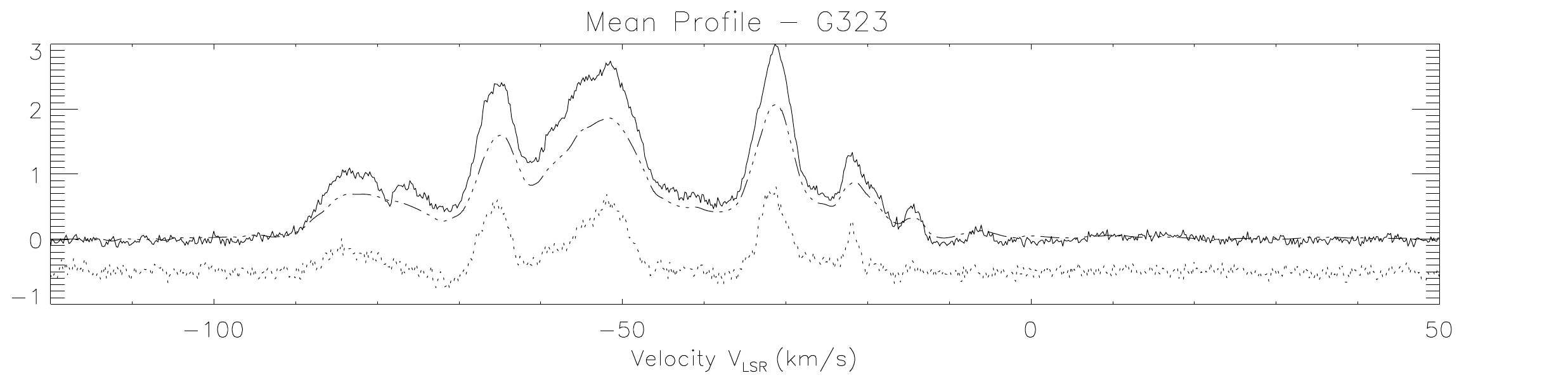}
\vspace{-3pt}
\includegraphics[height=\textwidth,angle=90]{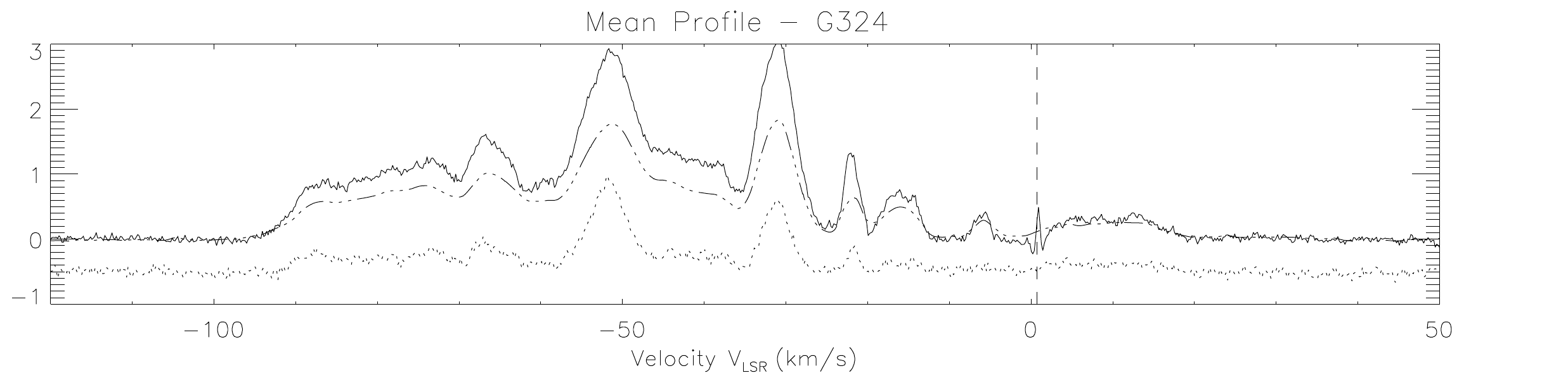}
\vspace{-21pt}
\caption{The CO line profiles, averaged over each square degree from $l = 320$--$325^\circ$, 
in units of $T_{\rm MB}$\,(K) (i.e.\ divided by the telescope efficiency, $\eta_{\rm XB} = 
0.55$). The solid line is $^{12}$CO; the dotted line is $^{13}$CO (which has been multiplied 
by 3 and offset by $-0.5$\,K for clarity); these spectra are binned in velocity by 2 pixels 
(0.2\,km\,s$^{-1}$). The dot-dashed line is the equivalent $^{12}$CO spectrum from the 
\citet{dht2001} survey, and the vertical dashed lines are the velocities from which 
contaminating sky emission was removed (see Table \ref{tab-sky}). 
\label{Fig6-Mean_Spectra}}
\end{center}
\end{figure*}

\begin{figure*}
\begin{center}
\includegraphics[height=\textwidth,angle=90]{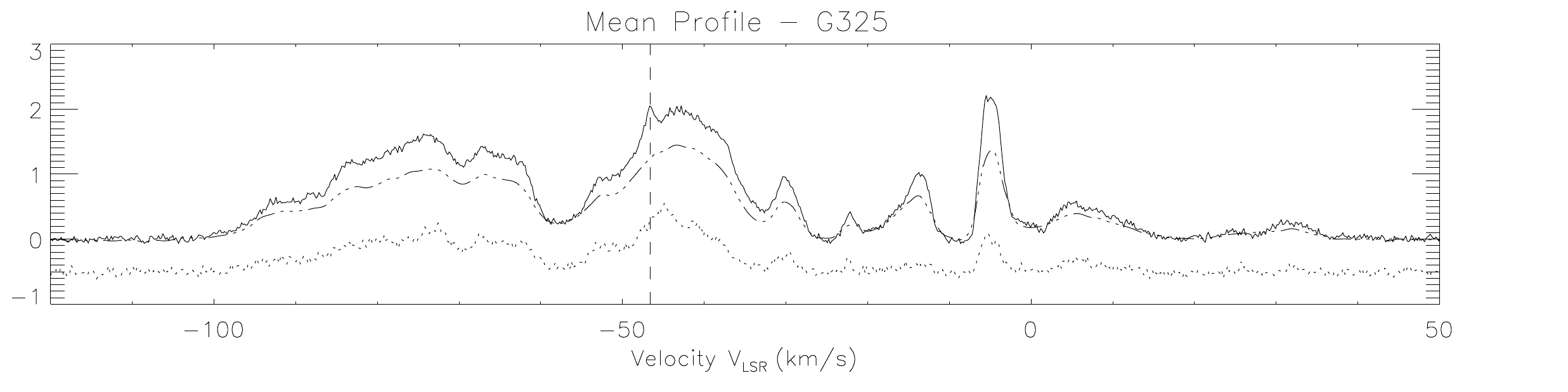}
\vspace{-3pt}
\includegraphics[height=\textwidth,angle=90]{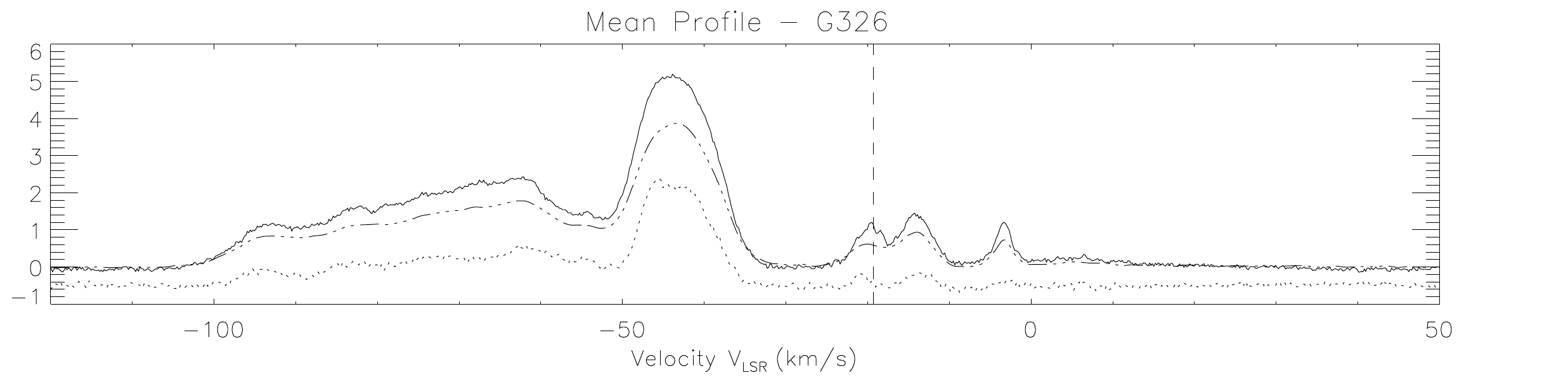}
\vspace{-3pt}
\includegraphics[height=\textwidth,angle=90]{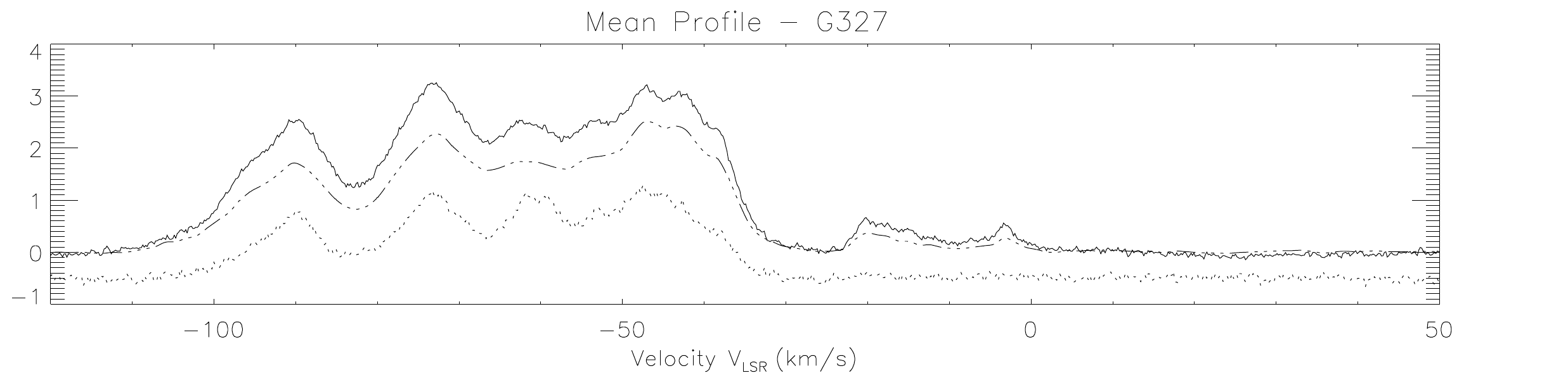}
\vspace{-3pt}
\includegraphics[height=\textwidth,angle=90]{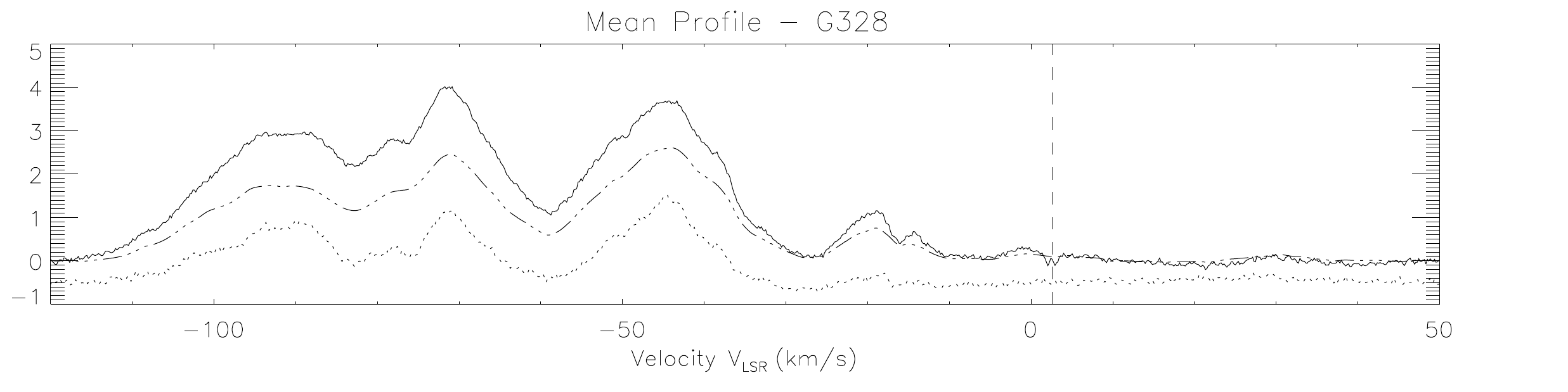}
\vspace{-3pt}
\includegraphics[height=\textwidth,angle=90]{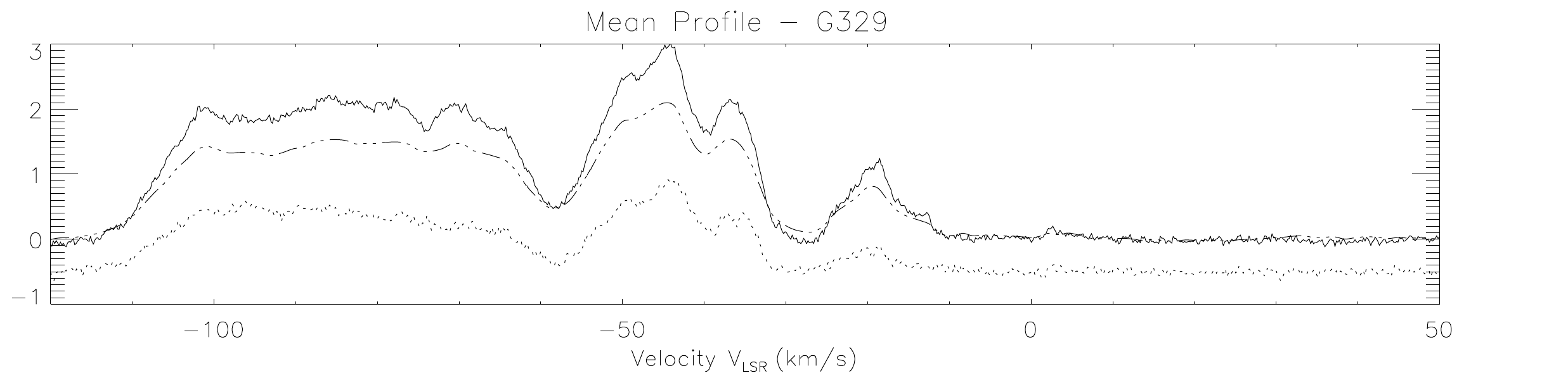}
\caption{Further CO line profiles, averaged over each square degree from $l = 
325$--$330^\circ$, in units of $T_{\rm MB}$\,(K). Other details are as in Figure 
\ref{Fig6-Mean_Spectra}. 
\label{Fig6-Mean_Spectra2}}
\end{center}
\end{figure*}

\begin{table}[t]
\caption{The median and mode of the ratio of the Mopra CO intensities to those of the 
\citet{dht2001} survey, calculated using the average spectra in each square degree over the 
velocity range $v = [-120,20]$\,km\,s$^{-1}$.}
\begin{center}
\begin{tabular}{@{}ccc@{}}
\hline\hline
Cube & Median & Mode  \\
\hline%
G320 & 1.42   & 1.40  \\
G321 & 1.31   & 1.27  \\
G322 & 1.28   & 1.28  \\
G323 & 1.42   & 1.37  \\
G324 & 1.62   & 1.61  \\
G325 & 1.41   & 1.37  \\
G326 & 1.34   & 1.31  \\
G327 & 1.38   & 1.27  \\
G328 & 1.64   & 1.60  \\
G329 & 1.41   & 1.37  \\
\hline
Average & 1.42 & 1.39 \\
\hline\hline
\end{tabular}
\end{center}
\label{tab-dht}
\end{table}
Figures \ref{Fig6-Mean_Spectra} and \ref{Fig6-Mean_Spectra2} show the average $^{12}$CO and 
$^{13}$CO line profiles for each square degree cube in DR1. The $^{12}$CO profiles from the 
lower spatial ($8'$) and spectral (1.3\,km\,s$^{-1}$) resolution survey of \citet{dht2001} are 
overlaid on the plots, showing that the structure of the profiles match well, although the 
intensity of the Mopra line peaks is $\sim1.4$ times higher than the corresponding 
\citet{dht2001} intensities (see Table \ref{tab-dht}). The intensity ratio was determined using 
the average spectra for each square degree over the velocity range $v = [-120,20]$\,km\,s$^{-1}$ 
where the majority of the emission lies, and has been found to be relatively consistent across 
DR1. \citet{wetal2011} found the average of the Mopra extended beam efficiency ($\eta_{\rm XB}$) 
had a typical rms deviation of $\sim$10\% over their eight observing seasons (each about a month 
in duration), with much of the variation reflecting changes in the system temperature scale due 
to instrument modifications. Their beam efficiencies were lower than those used to calculate 
$T_{\rm MB}$ in this paper; using these would only increase the ratio of the Mopra CO fluxes to 
those of \citet{dht2001}.

\begin{figure*}
\begin{center}
\includegraphics[height=\textwidth,angle=-90]{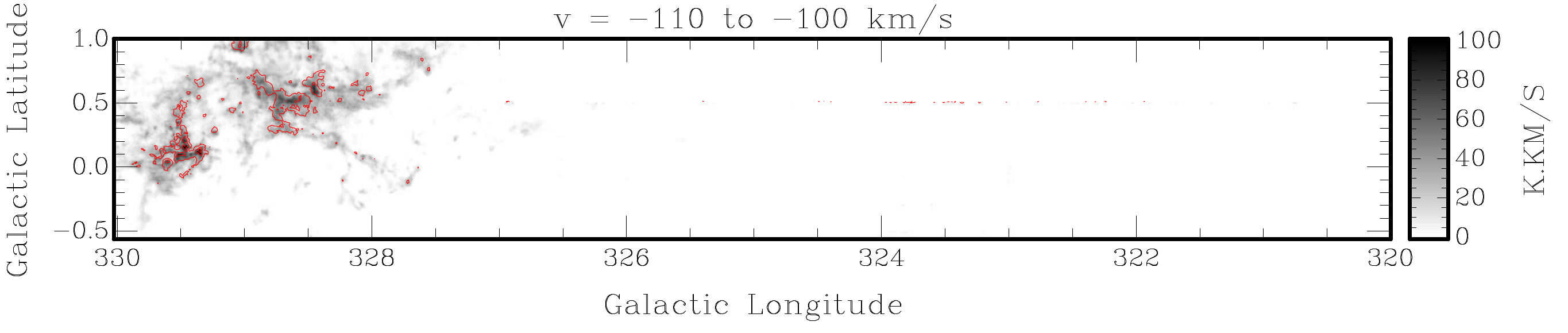}
\includegraphics[height=\textwidth,angle=-90]{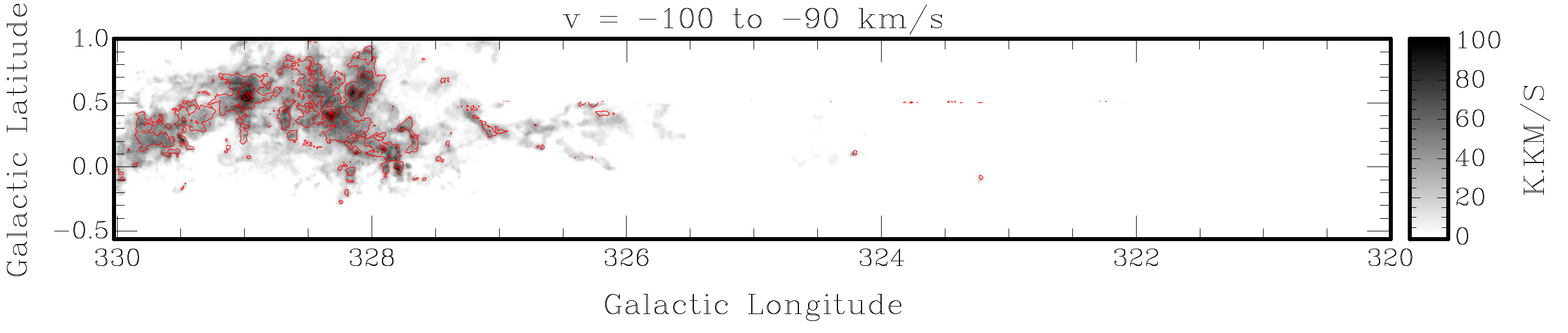}
\includegraphics[height=\textwidth,angle=-90]{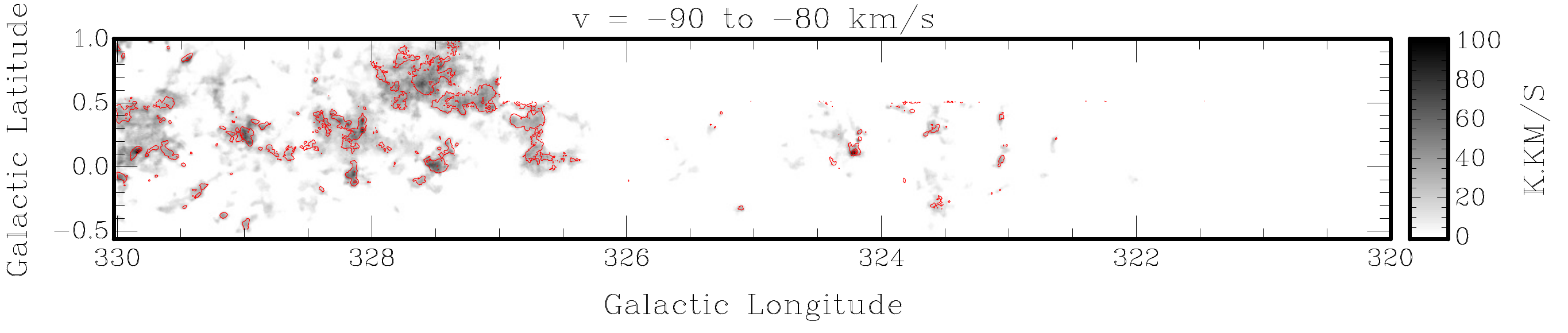}
\includegraphics[height=\textwidth,angle=-90]{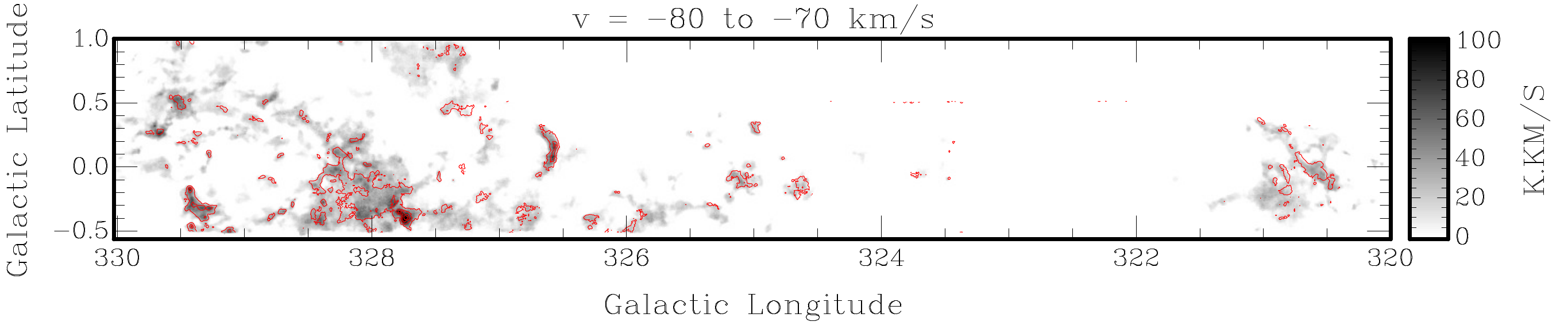}
\includegraphics[height=\textwidth,angle=-90]{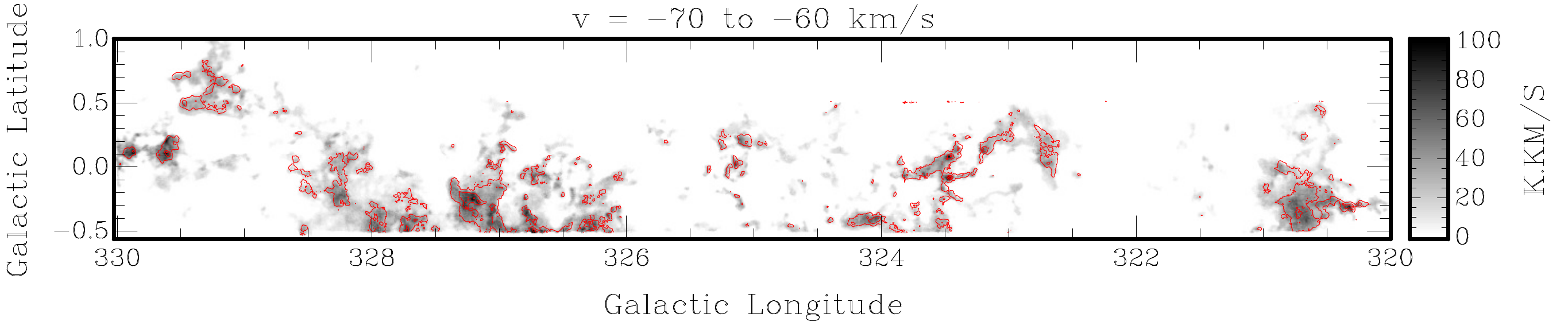}
\caption{$^{12}$CO intensity maps with $^{13}$CO contours in K\,km\,s$^{-1}$, corrected for the 
beam efficiency $\eta_\textrm{XB} = 0.55$, from $v = -110$ to $-60$\,km\,s$^{-1}$ integrated 
over 10\,km\,s$^{-1}$ intervals. The $^{13}$CO contours are at 5, 15, 25, 35 K\,km\,s$^{-1}$. 
Note that the boundary of the survey is evidenced by spots of poor signal to noise along the 
$b = + 0.5^\circ$ boundary.\label{Fig7-Moments}}
\end{center}
\end{figure*}

\begin{figure*}
\begin{center}
\includegraphics[height=\textwidth,angle=-90]{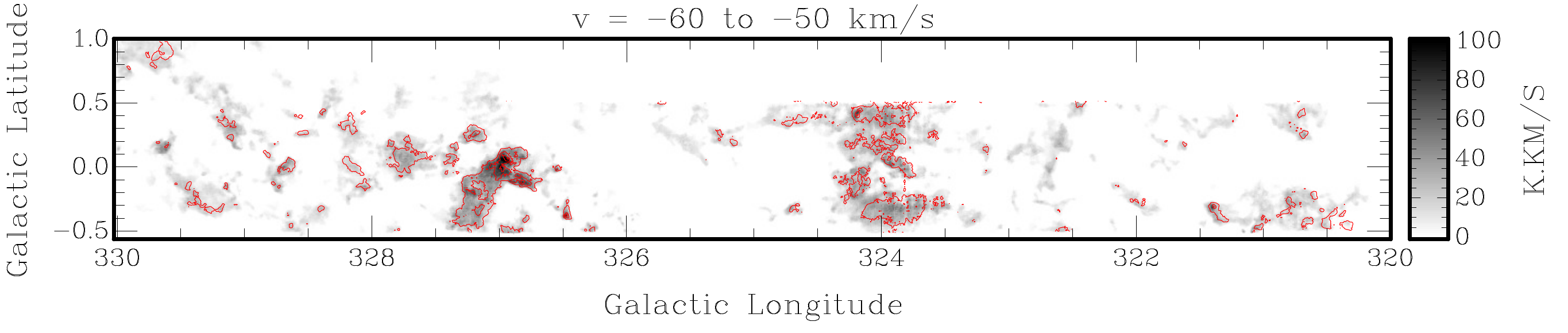}
\includegraphics[height=\textwidth,angle=-90]{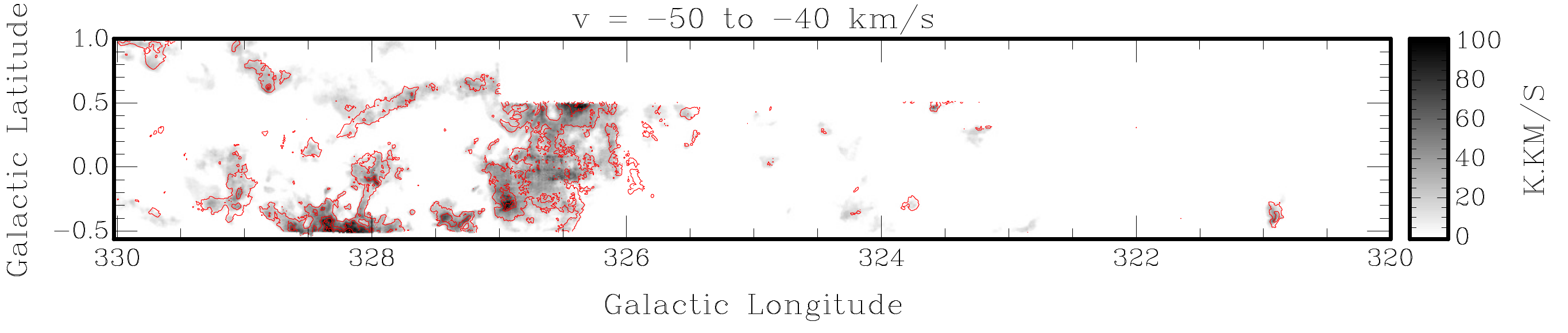}
\includegraphics[height=\textwidth,angle=-90]{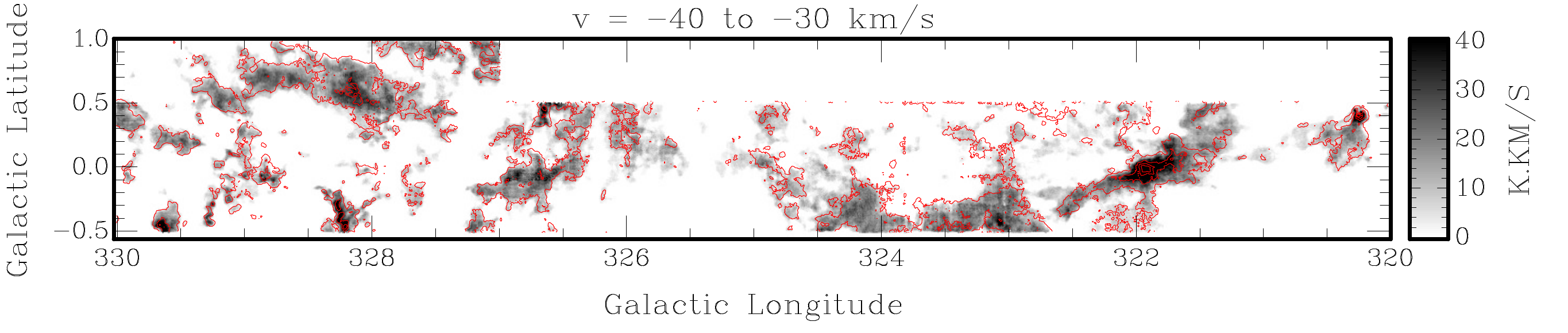}
\includegraphics[height=\textwidth,angle=-90]{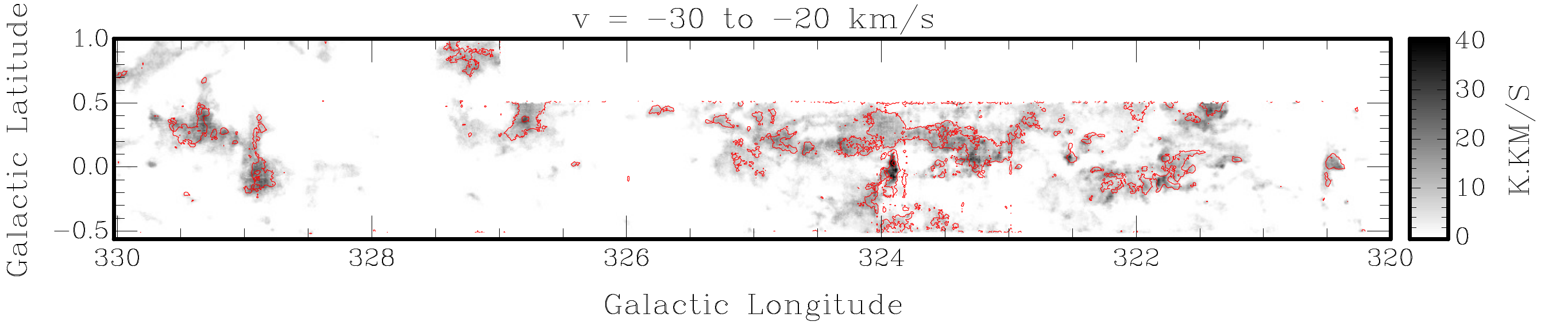}
\includegraphics[height=\textwidth,angle=-90]{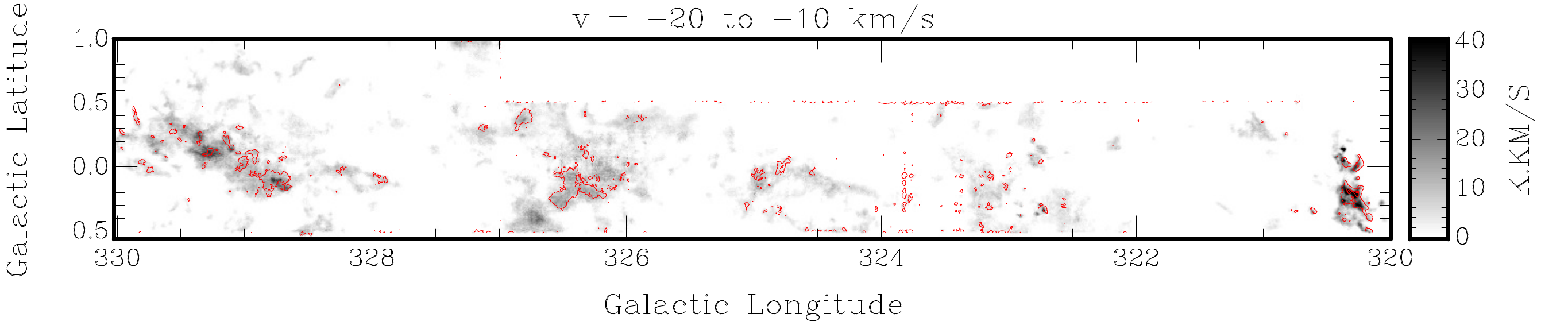}
\caption{As per Figure \ref{Fig7-Moments}, $^{12}$CO intensity maps with $^{13}$CO contours in 
K\,km\,s$^{-1}$, from $v = -60$ to $-10$\,km\,s$^{-1}$ integrated over 10\,km\,s$^{-1}$ intervals. 
Note that from the third map the intensity scale differs from Figure \ref{Fig7-Moments} due to 
the weaker fluxes in these velocity ranges, and the $^{13}$CO contours are now at 1, 8, 15, 
22 K\,km\,s$^{-1}$. Due to the weaker signal fluxes there are more localised instances of high 
noise apparent in the $^{13}$CO contours in this Figure, most obviously the column at $l \approx 
323.8^\circ$ in the lower three maps. \label{Fig7b-Moments}}
\end{center}
\end{figure*}

\begin{figure*}
\begin{center}
\includegraphics[height=\textwidth,angle=-90]{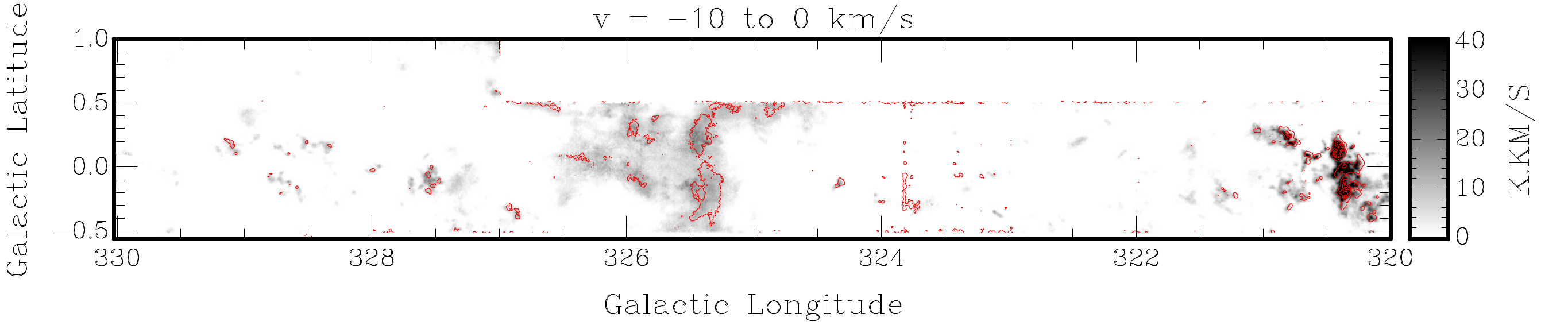}
\includegraphics[height=\textwidth,angle=-90]{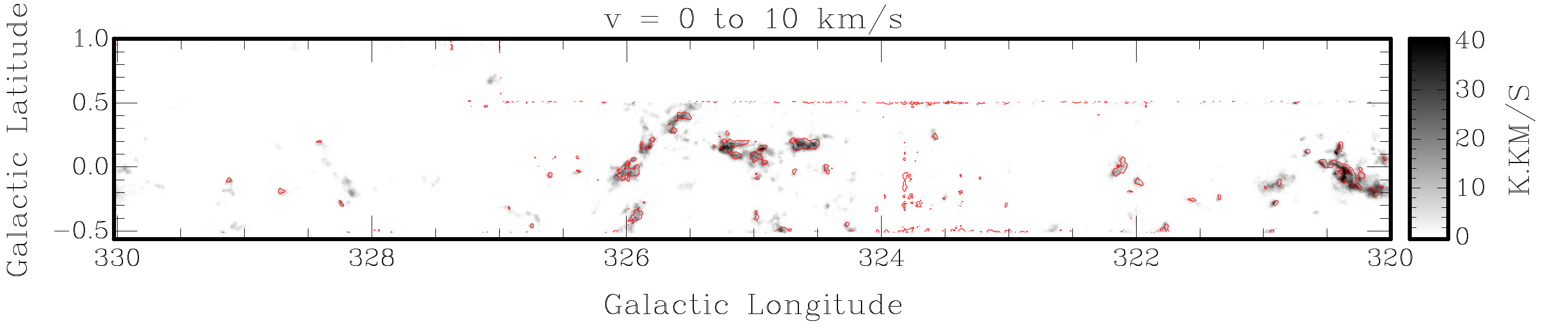}
\includegraphics[height=\textwidth,angle=-90]{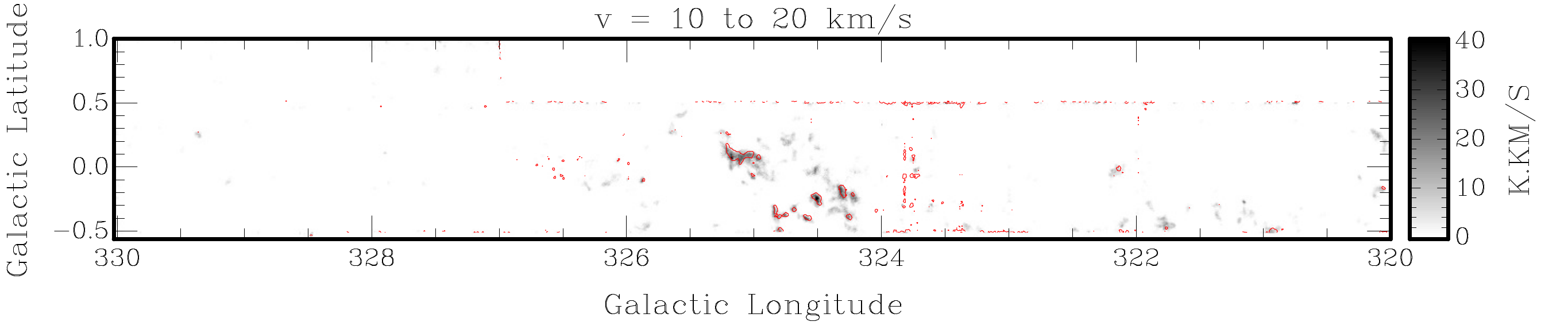}
\includegraphics[height=\textwidth,angle=-90]{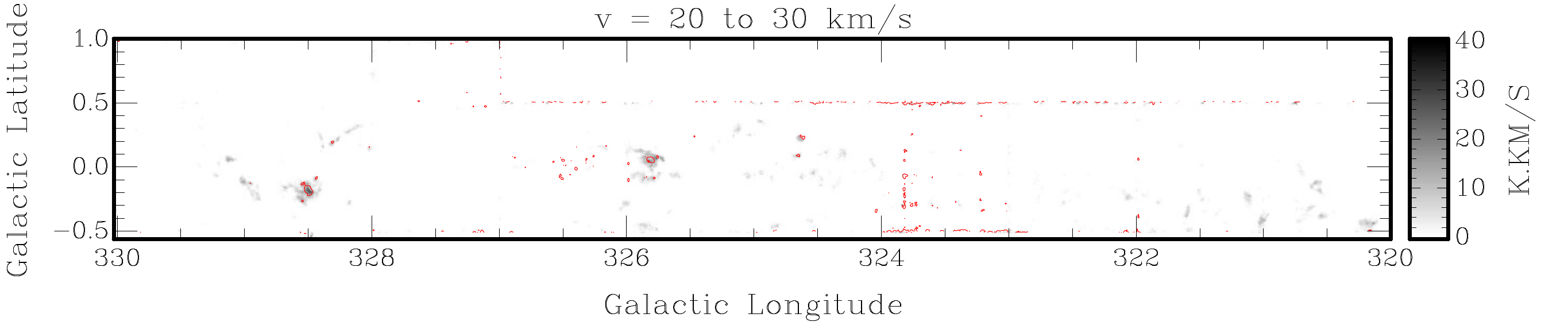}
\includegraphics[height=\textwidth,angle=-90]{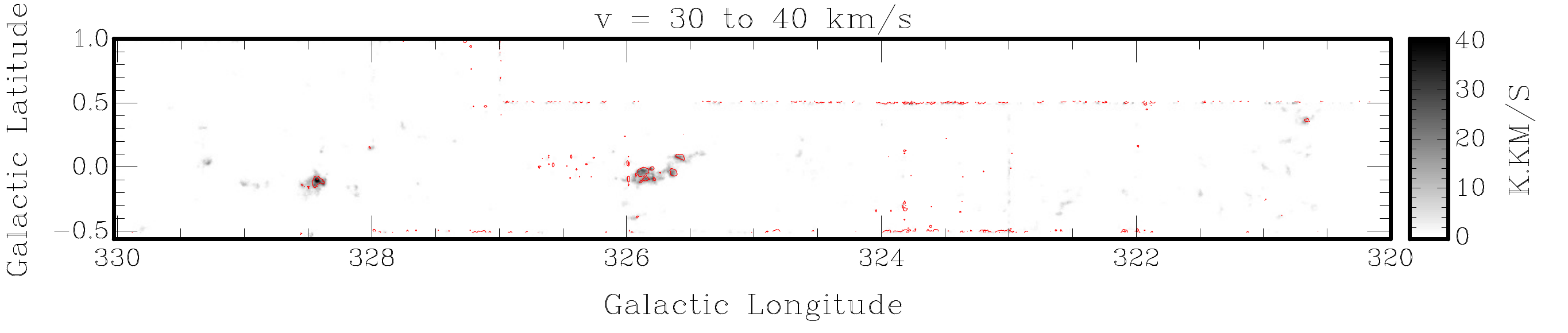}
\caption{As per Figure \ref{Fig7b-Moments}, $^{12}$CO intensity maps with $^{13}$CO contours in 
K\,km\,s$^{-1}$, from $v = -10$ to $+40$\,km\,s$^{-1}$ integrated over 10\,km\,s$^{-1}$ intervals. 
\label{Fig7c-Moments}}
\end{center}
\end{figure*}

The integrated intensity maps in Figures \ref{Fig7-Moments}--\ref{Fig7c-Moments} highlight 
molecular clouds in the Mopra $^{12}$CO and $^{13}$CO data cubes over each 10\,km\,s$^{-1}$ 
interval between $-110$ and $+40$\,km\,s$^{-1}$ in units of K\,km\,s$^{-1}$, corrected for 
the beam efficiency $\eta_\textrm{XB} = 0.55$. These have been calculated using the 
\textit{dilated mask} code developed for the MAGMA survey of the Magellanic Clouds in CO 
\citep{wetal2011}. The dilated CPROPS mask \citep{rl2006} is calculated by identifying 
regions of high significance ($> 4\sigma$) and expanding these to connected regions of lower 
significance ($> 3\sigma$). Masking the data in this way reduces the number of high noise 
artefacts in the intensity images, although some remain. In particular, the upper boundary of 
the main survey at $l = 320$--$327^\circ$, $b = +0.5^\circ$ is clearly visible as a series of 
high noise points in both the $^{12}$CO intensity map and the $^{13}$CO contours. 

\begin{figure*}
\begin{center}
\includegraphics[width=350pt]{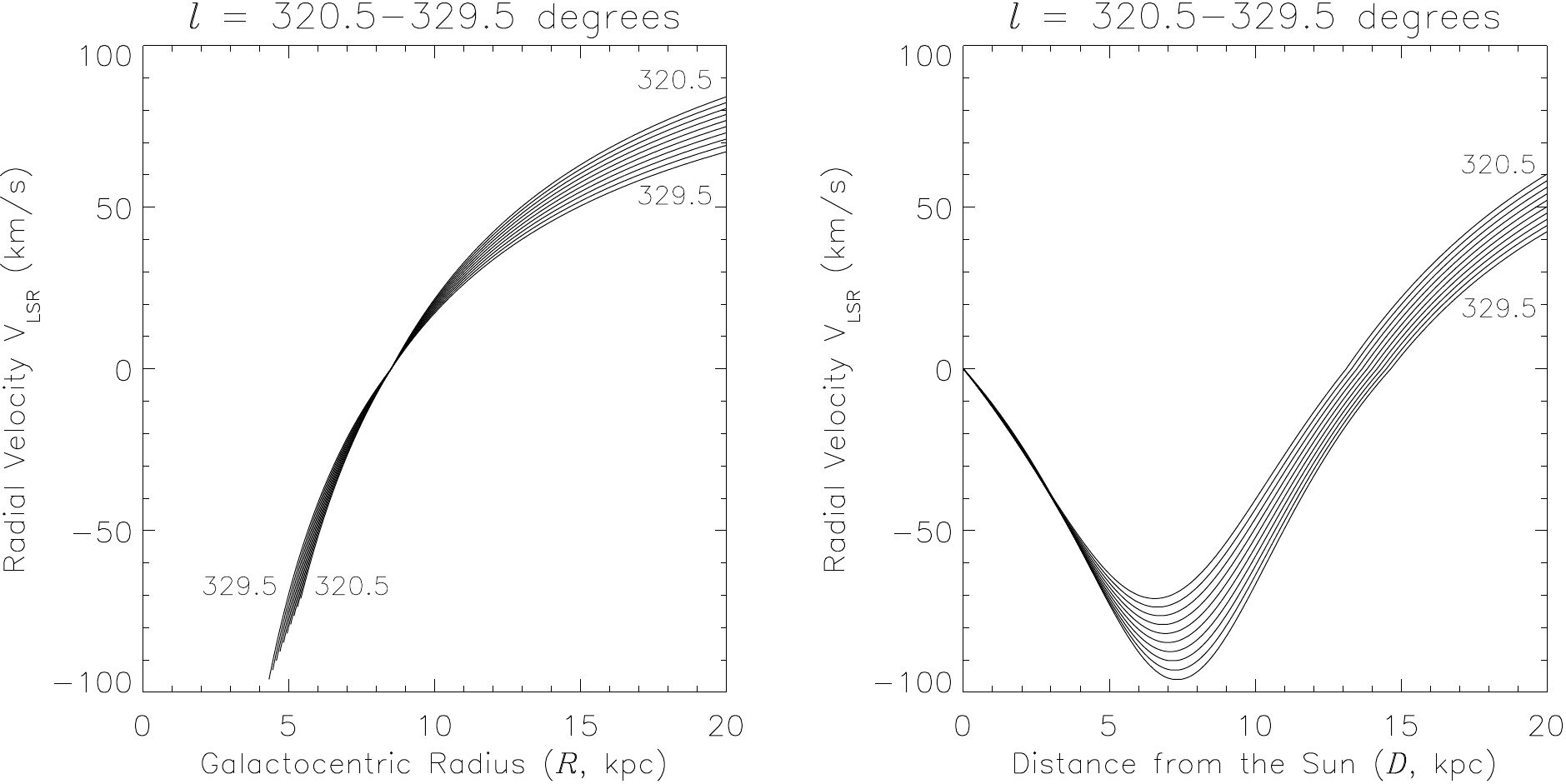}
\caption{Radial velocity-distance relationships calculated for the centre of each square degree 
in DR1 using the \citet{mgd2007} rotation curve for the inner Galaxy (i.e.\ negative velocities, 
with $R < R_\odot$) in the fourth quadrant, with the \citet{bb1993} curve for the outer Galaxy 
(i.e.\ positive velocities, and scaled to give the same orbital velocity at $R = R_\odot$). To 
the left, the galactocentric radius in kpc is plotted against radial velocity, $V_{\rm LSR}$ in 
km\,s$^{-1}$. To the right, distance from the Sun, $D$, in kpc is plotted against $V_{\rm LSR}$. 
Near-distance solutions assume $D < D_{\rm tangent} \sim 7.0$\,kpc. Far-distance solutions are 
for $D_{\rm tangent} < D < 2 D_{\rm tangent}$. 
\label{Fig5-RotationCurves}}
\end{center}
\end{figure*}
To aid interpretation of the data, Figure \ref{Fig5-RotationCurves} shows derived relationships 
between the radial velocity and galactocentric radius ($R$, left) and the distance from the Sun 
($D$, right) for the central longitude of each square degree in DR1. This is an expansion of 
Fig.\ $9$ in Paper I, now showing the variation across the whole ten degrees. As in that figure, 
the right plot shows the near-far ambiguity for negative velocities between 0\,km\,s$^{-1}$ and 
the tangent velocity along each sightline (decreasing with $l$). From the intensity maps in 
Figure \ref{Fig7-Moments}, it is clear that a number of molecular clouds occur at `forbidden' 
velocities (possessing velocities that are not possible under the assumption that the rotation 
curve holds, i.e.\ $< -95$\,km\,s$^{-1}$); in this analysis any such velocities are assigned to 
the tangential distance. Clouds with positive velocities have source distances greater than 
twice the tangent distance. 

\begin{figure*}
\begin{center}
\includegraphics[height=\textwidth,angle=90]{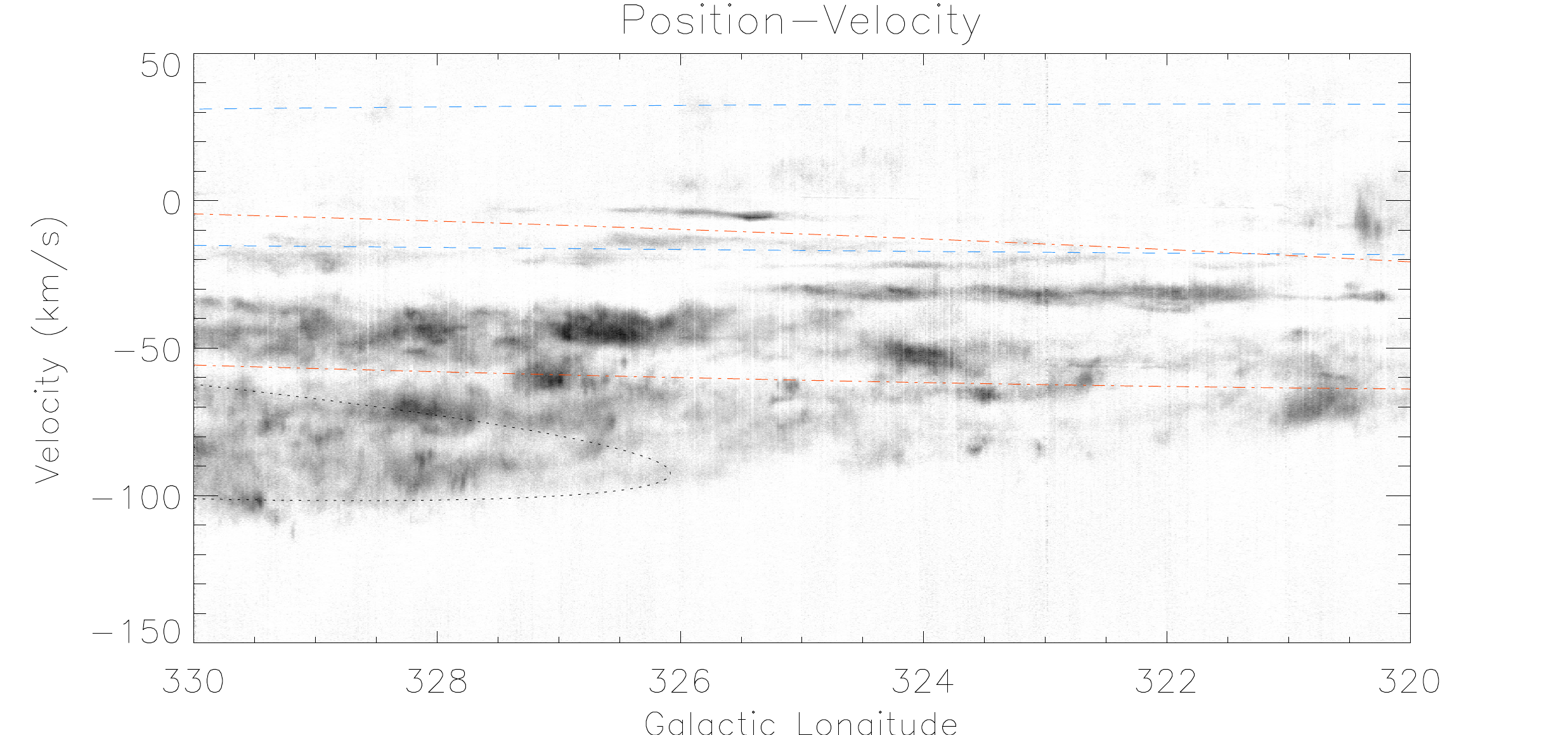}
\caption{$^{12}$CO position-velocity image for DR1, with the Galactic longitude, $l$ on the 
$x$-axis, against the $V_{\rm LSR}$ radial velocity in km\,s$^{-1}$ on the $y$-axis. The data 
have been averaged over the central degree in latitude; residuals from poor data are 
evident in the vertical striping. The solid lines are model positions of the centre of the 
spiral arms in a four-arm spiral galaxy with pitch angle of $12.5^\circ$, a central bar length 
of 3\,kpc and a Sun-Galactic centre distance of 8.0\,kpc \citep[parameters from][]{v2014}. From 
the lower edge on the left axis these are: the Norma arm (near at $-100$\,km\,s$^{-1}$ and far at 
$-60$\,km\,s$^{-1}$; black dotted line), the Scutum-Crux arm (near; red dot-dashed line), the 
Sagittarius-Carina arm (near; blue dashed line), the Scutum-Crux arm (far; red dot-dashed line) 
and the Sagittarius-Carina arm (far; blue dashed line). Only a few clouds are visible in this 
furthest arm, and only in $^{12}$CO and $^{13}$CO. \label{Fig8-PVplot}}
\end{center}
\end{figure*}

Figure \ref{Fig8-PVplot} shows a position-velocity slice of the $^{12}$CO emission averaged 
over the primary degree of latitude of the DR1 data cube, overlaid with model positions of 
the spiral arms in a four-arm spiral galaxy. The parameters of the model galaxy have been 
taken from \citet{v2014} and depict a four-arm spiral galaxy with a pitch angle of 
$12.5^\circ$, a central bar length of 3\,kpc and Sun-Galactic centre distance of 8.0\,kpc. 
While the molecular clouds clearly lie in striped bands across the position-velocity diagram, 
the exact centre of the spiral arms is harder to determine. The meta-analysis of \citet{v2014} 
calculated the radius of the spiral arms to be near $400$\,pc from mid-arm to dust lane, which 
suggests that it is difficult to identify clearly the arm in which individual clouds reside, 
particularly in regions such as $l = 329$--$330^\circ$, where the Norma and Scutum-Crux arms are 
close in velocity space. 

\begin{figure}
\begin{center}
\includegraphics[height=\columnwidth,angle=-90]{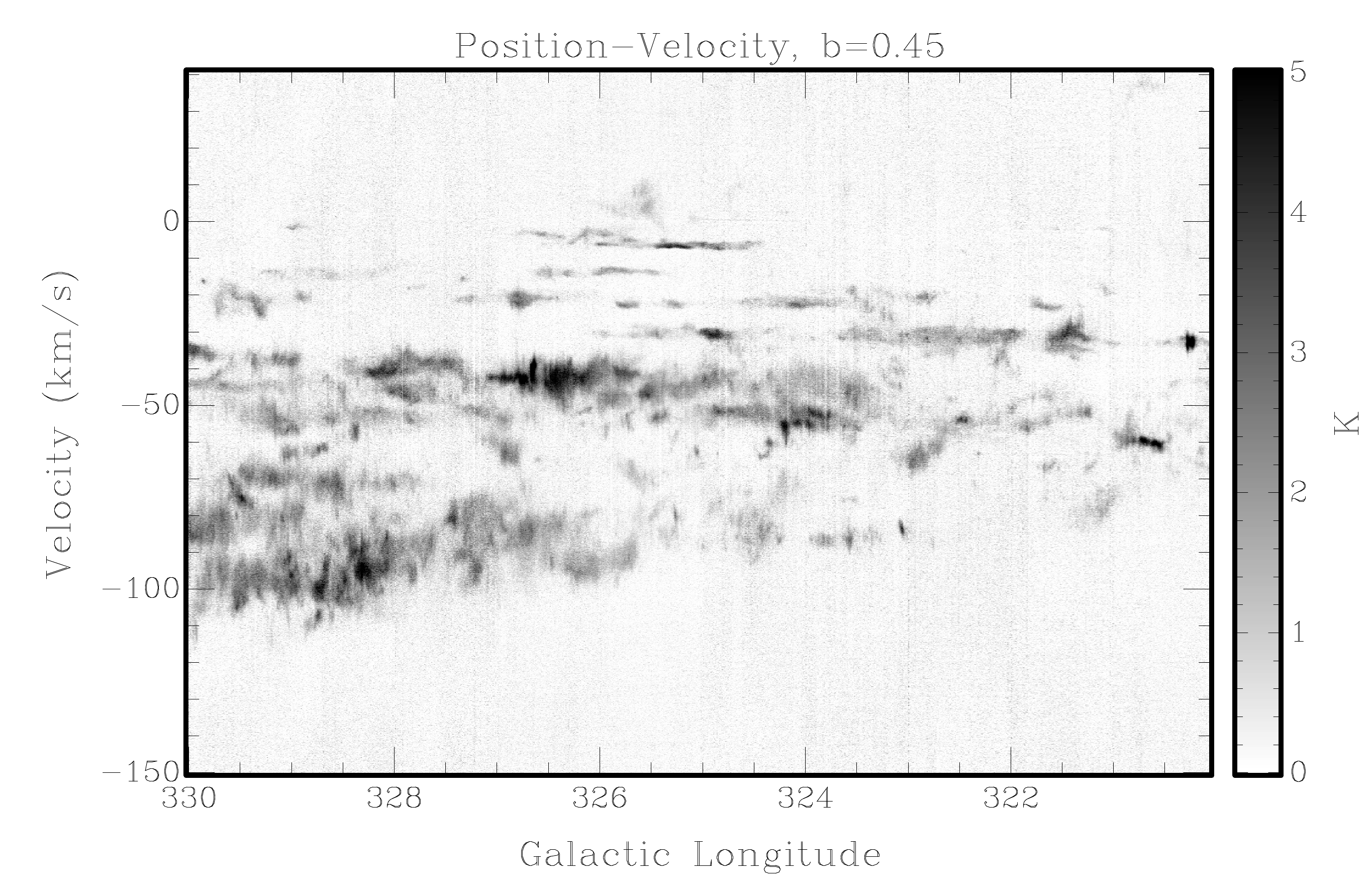}
\includegraphics[height=\columnwidth,angle=-90]{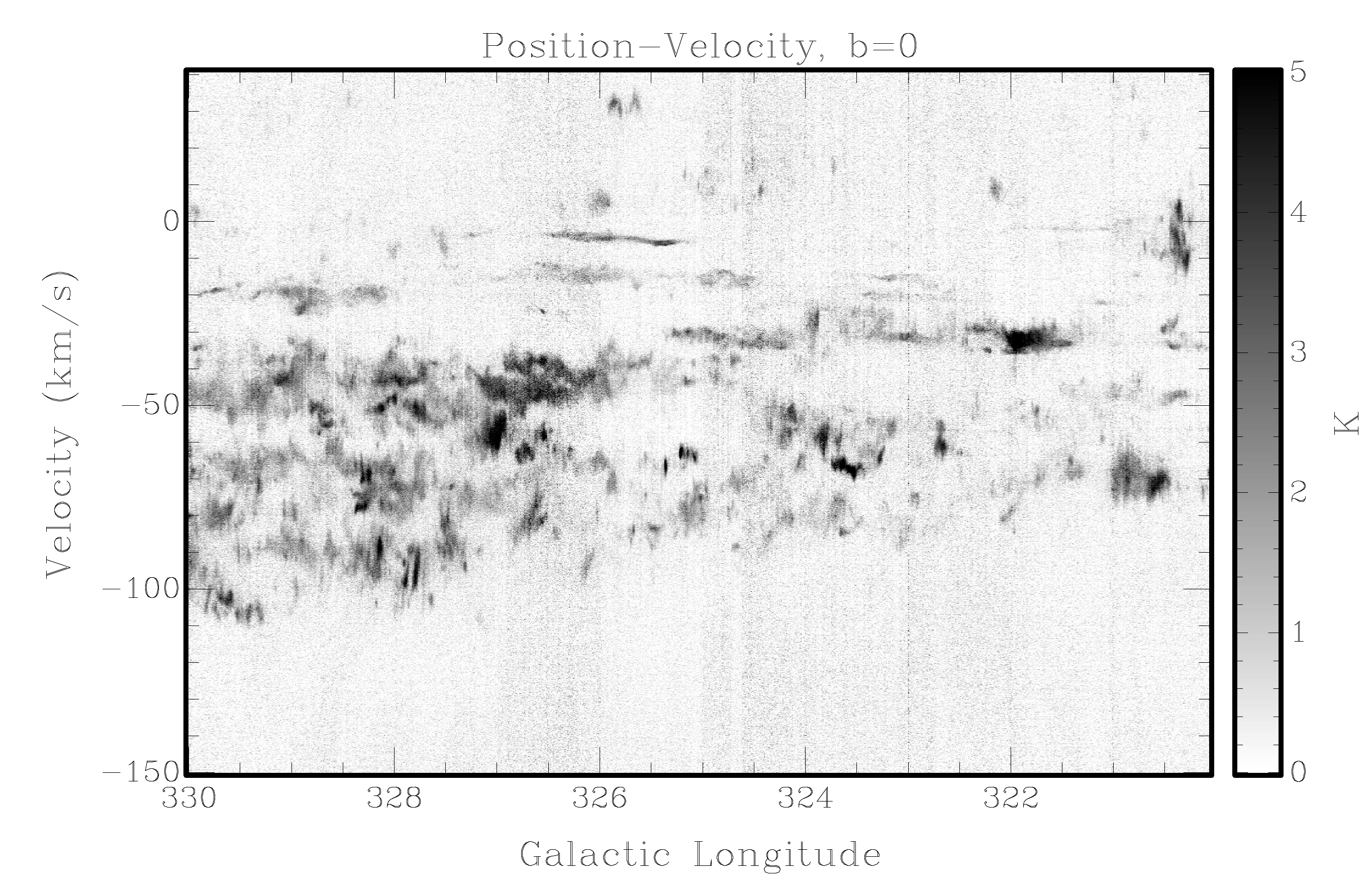}
\includegraphics[height=\columnwidth,angle=-90]{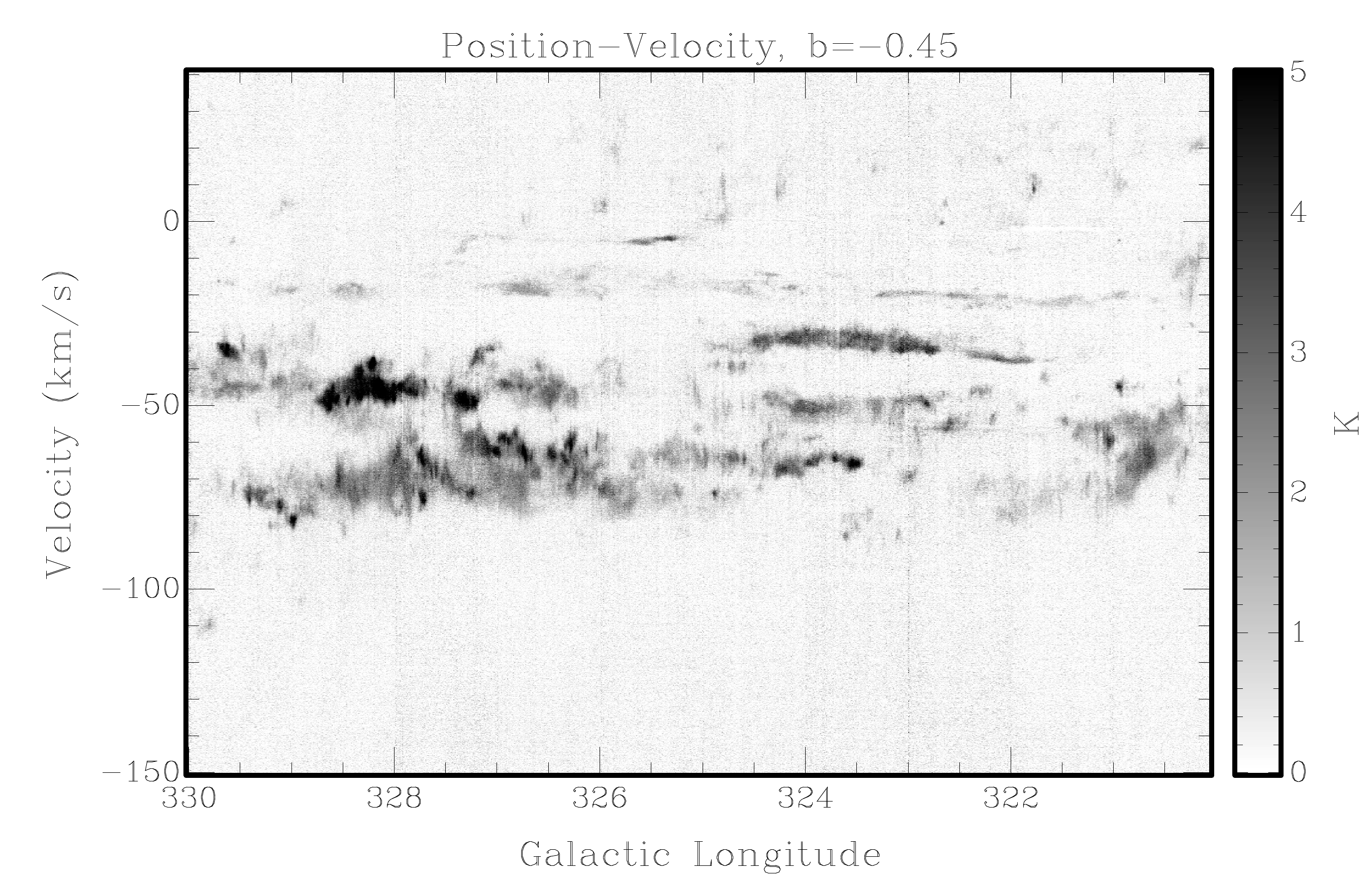}
\caption{$^{12}$CO position-velocity images for DR1, with the Galactic longitude, $l$ on the 
$x$-axis, against the $V_{\rm LSR}$ radial velocity in km\,s$^{-1}$ on the $y$-axis. The data 
in each plot have been averaged over 6 arcmin in latitude, the upper image is centred on $b = 
0.45^\circ$, the central plot is the midplane $b=0^\circ$ and the lower image is centred on 
$b = -0.45^\circ$; residuals from poor data are evident in the vertical striping. 
\label{Fig8b-ExtraPV}}
\end{center}
\end{figure}
The distribution of molecular clouds with latitude is illustrated in Figure \ref{Fig8b-ExtraPV}, 
which shows the position-velocity distribution of $^{12}$CO emission vertically averaged over the 
6 arcmin surrounding each of $b=0^\circ$ and $b = \pm 0.45^\circ$. The differences between these 
images is striking, showing how discrete the molecular gas is even within the Galactic plane. 
There is very little CO located at the `forbidden' velocities indicated by Figure 
\ref{Fig5-RotationCurves} at $b = -0.45^\circ$, while the fastest negative velocities occur at 
the higher latitudes. It is also worth noting that the vertical streaking, caused by high noise 
residuals in the data, occurs at different longitudes in the three position-velocity diagrams; 
these suggest that the methods used to clean high-noise columns work well, although individual 
pixels with poor baselines may remain. 

\begin{figure}
\begin{center}
\includegraphics[width=\columnwidth]{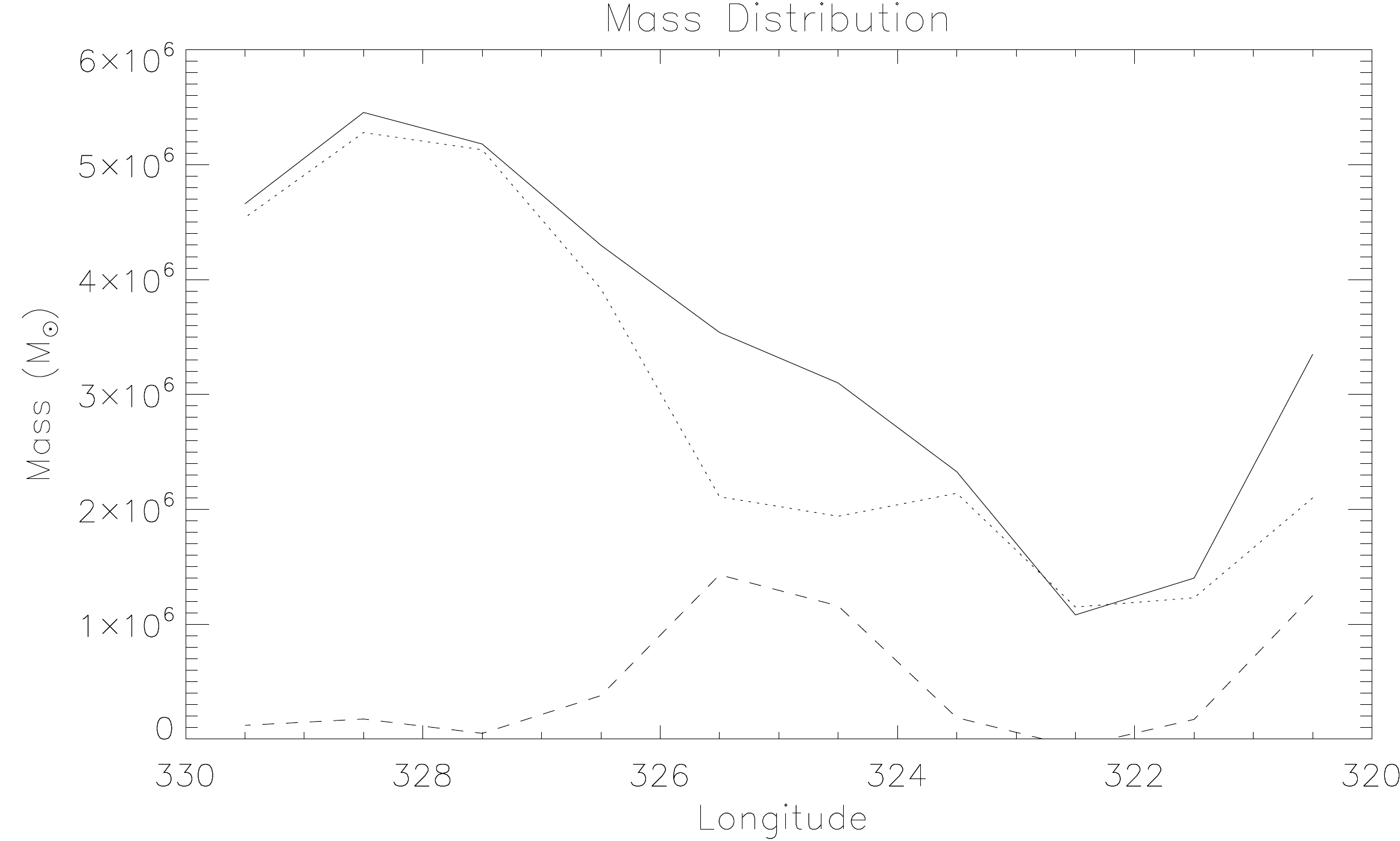}
\caption{Mass distribution in DR1, calculated from the $^{12}$CO emission per square degree 
using X$_{\rm CO} = 2.7 \times 10^{20}$\,cm$^{-2}$\,(K\,km\,s$^{-1}$)$^{-1}$ (see text). 
The dotted line is the mass contribution from molecular clouds at negative velocities (assumed 
to be at the near distance), the dashed line that from positive velocities (sited beyond the 
Solar circle), and the solid line is the total (all in units of Solar mass).\label{Masses}}
\vspace{-10pt}
\end{center}
\end{figure}

Figure \ref{Masses} shows the distribution of mass in DR1, under the assumption that the clouds
with negative velocities are located at the near distance (providing a lower limit on the true
molecular mass), and that the X-factor used to link the CO intensity to the H$_2$ column density 
is X$_{\rm CO} = 2.7 \times 10^{20}$\,cm$^{-2}$\,(K\,km\,s$^{-1}$)$^{-1}$. This value is 
adopted for $|b|<1^\circ$ using figure 11 of \citet{dht2001}, and while it is larger than 
their value of X$_{\rm CO} = 1.8 \times 10^{20}$\,cm$^{-2}$\,(K\,km\,s$^{-1}$)$^{-1}$ for 
$|b|>5^\circ$, it is within the uncertainties recommended by \citet{bwl2013} in their 
meta-analysis of X-factor values from observational and theoretical studies of the Milky Way. 
These simplifications were also adopted in Paper I to provide a lower limit on the molecular 
mass in G323; however it was also shown that one of the five features in 
the spectrum was likely mis-identified as being at the near distance rather than the far 
distance. This leads to an underestimate of the total mass of about one quarter, which is within 
the uncertainty of using a constant X-factor mass calibration. 

The mass distribution in Figure \ref{Masses}, while coarse, shows how the lower mass 
limit for each square degree varies over these $10^\circ$ in longitude. The contributions 
to the total mass per degree from clouds at positive (far distances) and negative (nearer) 
velocities can be related to features in the position-velocity diagram in Figure 
\ref{Fig8-PVplot}. In particular, the lines of sight through $l = 324$--$326^\circ$ possess 
a large component of mass at 
positive velocities, arising from the clouds at $+35$\,km\,s$^{-1}$ from the far 
Sagittarius-Carina arm (see Figure \ref{Fig7c-Moments}), while the increase at $l = 320^\circ$ 
is due to a large cloud at $v = [-10,+10]$\,km\,s$^{-1}$, which cannot be associated with a 
single spiral arm using only the CO emission. The mass increases from a minimum at $l \sim 
322^\circ$ of $\sim10^6$\,M$_\odot$ to a maximum at $l \sim 328^\circ$ of $\sim5 \times 
10^6$\,M$_\odot$, with a total of $\sim4 \times 10^7$\,M$_\odot$ within the $10^\circ$ main 
survey span, where $|b|<0.5^\circ$. Of this, approximately 15\% is situated beyond the Solar 
circle.

\subsection{Sample Science}

The principal intention of this paper is to describe the CO data cubes that make up DR1.  These 
may be used for a variety of applications, as they provide the distribution of the molecular gas 
along a substantive portion of the fourth quadrant of the Galaxy, as well as its dynamical 
motions.  From this, column densities, optical depths and the three dimensional molecular mass 
distribution can be inferred, such as were outlined above.  Comparison with infrared to 
millimetre-wave continuum images, for instance, would allow star formation efficiencies to be 
determined, and the variation of the X$_{\rm CO}$ conversion factor between CO flux and H$_2$ 
column density to be examined on a galactic scale.  Three sample science investigations that 
have been undertaken using the Mopra CO survey data are illustrated here.

The first compares the CO emission in the G328 region with that of [CI] 809\,GHz (from the HEAT 
telescope in Antarctica) and HI 21\,cm \citep[from the Southern Galactic Plane 
Survey;][]{mgetal2005} emission, discovering a cold, filamentary molecular cloud nearly 
$1^{\circ}$ in length but only $5'$ across \citep{betal2014}.  The emission was confined to a 
narrow (2\,km\,s$^{-1}$ FWHM) velocity range, with the hydrogen seen in self-absorption (i.e.\ 
HISA), apparently enveloping the molecular gas.  The authors hypothesised that the formation of 
a molecular cloud was being witnessed, with the molecular material possibly condensing out of 
the atomic substrate.  Further analysis of the entire data cube, not just the narrow velocity 
range associated with this filament, is now being undertaken to investigate the variation of the 
[CI] to CO ratio, and to search for evidence of `dark' molecular gas.  This is molecular gas 
where its normal tracer (CO) is weak, or absent (i.e.\ H$_2$ without CO), so that the molecular 
gas has not been detected, or at least the amount has been under-estimated. However if CO is 
absent (presumably due to photodissociation by far--UV photons), the carbon will be present as 
either C or C$^+$.  These species emit in the THz spectral regime, where the HEAT telescope 
operates.  Analysis of the complete data set (Burton et al. 2015, in preparation) indicates that 
at the edges of the molecular clouds there is a tendency for the [CI]/CO ratio to rise, as the CO 
itself falls, consistent with the existence of dark molecular gas.

The second example relates the Mopra CO data to X-ray emission.  A bright X-ray flare was 
observed from the binary star Circinus X--1 in late 2013 by three spacecraft (\textit{Swift, 
Chandra} and \textit{XMM-Newton}).  Over a three month period four well-defined rings were seen 
in the X-rays, extending outwards with time to a distance of $14'$ from the central source. This 
is in the G322 portion of our CO survey. Comparison of the Mopra data with the images from the 
X-ray satellites showed that each of the X-ray peaks in the rings is correlated with 
corresponding emission in the molecular gas.  However, each has a different velocity, ranging 
from $-81$ to $-32$\,km\,s$^{-1}$ \citep[see][]{hetal2015}.  The X-rays are being scattered off 
clumps of dust associated with the molecular gas to produce the light echoes.  Using the 
distances to these clumps inferred from the Galactic rotation curve, and the scattering geometry 
a direct kinematic distance could be determined for the source.  This was found to be 9.4\,kpc, 
significantly greater than the previously estimated 4\,kpc.  It implies that Circinus X--1 is a 
frequent super-Eddington source, producing a narrow jet with a collimation angle of $\theta 
< 3^{\circ}$ and a Lorentz beaming factor of $\Gamma > 22$.

The final example relates the Mopra CO data to an unidentified radio continuum source known as 
PMN 1452-5910, thought possibly to be related to an SNR evident at 843\,MHz from MOST data 
\citep{bls1999}, as analysed by \citet{jb2015}.  Three features are apparent in the Mopra CO 
spectrum towards this source, at $-50, -42$ and $-1$\,km\,s$^{-1}$. Their morphology, however, 
in comparison to that of the radio continuum, suggests that it is the $-1$\,km\,s$^{-1}$ CO 
feature that is associated with the radio, so providing a distance to the source.  This also 
identifies the source with an H$_2$O maser at a similar velocity, and enables an interpretation 
of infrared images from the Spitzer telescope to be made.  Jones \& Braiding concluded that the 
source is a previously-unidentified massive star forming region about 13\,kpc away.  They then 
determine parameters for the region: $\sim 2 \times 10^4\,$M$_\odot$ of molecular gas with an 
average column of $\sim 10^{21}\,$cm$^{-2}$, enclosing several UCH\textsc{ii} regions with 
emission measures $\sim 10^{6-7}$\,pc\,cm$^{-6}$ and electron densities $\sim 10^3$\,cm$^{-3}$.  
Three OB stars (2 of type B0, 1 of O8) can account for the radio emission.

\section{CONCLUSIONS}

In their forthcoming Annual Review of molecular clouds in the Milky Way, \citet{hd2015} state: 
``For further progress in [molecular cloud census studies], surveys comparable to the Galactic 
Ring Survey \citep{grs} in resolution and sensitivity in the fourth quadrant are required.'' The 
first $10^\circ$ of such a survey have been presented here. The Cherenkov Telescope Array will 
produce a new survey of the high energy gamma-ray sources in the Galactic plane \citep{detal2013} 
at a resolution approaching 1 arcminute \citep{bernloehr2013}. High resolution molecular gas 
surveys such as this will be required to properly identify these sources amongst the diffuse TeV 
gamma-ray emission components CTA will likely detect \citep[e.g.][]{acero2013}. The Mopra CO 
survey of the southern Galactic plane aims to address these community needs by observing the 
entirety of the fourth quadrant along the plane from $l = 270$--$360^\circ$ where $|b| < 
0.5^\circ$, at a spatial resolution of 35 arcsec and a spectral resolution of 0.1\,km\,s$^{-1}$, 
in the $J = 1$--0 lines of $^{12}$CO, $^{13}$CO, C$^{18}$O and C$^{17}$O. This is a significant 
improvement over previous studies conducted of the fourth quadrant of the Galaxy. 

Data Release 1 from the Mopra CO survey covers the 11.5 square degrees from $l = 
320$--$330^\circ$, $b = \pm 0.5^\circ$ and $l = 327$--$330^\circ$, $b = +0.5$--$1.0^\circ$. 
Included in this paper are a number of metrics describing the quality of the dataset, which has 
a typical rms noise of $\sim 1.3$\,K in the $^{12}$CO frequency band and $\sim 0.5$\,K in the 
other isotopologue frequency bands, and a discussion of the improvement in the data taken during 
the winter months compared to the initial square degree observed in summer that was published in 
Paper I. The line intensities in the Mopra CO survey are $\sim 1.4$ times those of the 
\citet{dht2001} survey, and our higher resolution CO line profiles match those of their survey 
quite well. Sample line profiles for each square degree are presented, as well as integrated 
intensity maps that show the filamentary nature and complexity of the molecular clouds in the 
survey. The primary $10^\circ \times 1^\circ$ strip of the Galactic plane in DR1 has been found 
to contain $\sim4 \times 10^7$\,M$_\odot$ of molecular gas.

The Mopra CO survey is ongoing, with the intention of covering the entirety of the fourth 
quadrant of the Galactic plane by the end of 2015. The data are being made available at the 
survey website\footnote{http://www.phys.unsw.edu.au/mopraco/}, and in the CSIRO-ATNF data 
archive\footnote{http://atoa.atnf.csiro.au}, as they are published. 

\begin{acknowledgements}
The Mopra radio telescope is currently part of the Australia Telescope National Facility. 
Operations support was provided by the University of New South Wales and the University of 
Adelaide. Many staff of the ATNF have contributed to the success of the remote operations 
at Mopra.  We particularly wish to acknowledge the contributions of David Brodrick, Philip 
Edwards, Brett Hiscock, Balt Indermuehle and Peter Mirtschin. The University of New South Wales 
Digital Filter Bank used for the observations with the Mopra Telescope (the UNSW--MOPS) was 
provided with support from the Australian Research Council (ARC). We also acknowledge ARC 
support through Discovery Project DP120101585.   

Finally, we thank the anonymous referee whose thoughtful comments have improved the 
clarity and accessibility of this paper.
\end{acknowledgements}


\end{document}